\documentclass[%
 reprint,
 amsmath,amssymb,
 aps,
]{revtex4-2}

\usepackage{amsmath}
\usepackage{enumitem}
\usepackage{graphicx}
\usepackage{dcolumn}
\usepackage{bm}

\begin{document}

\preprint{APS/123-QED}

\title{Identifying $D_{sJ}^*(2860)$ as four resonance states\\ through strong decay analysis}

\author{Keval Gandhi}
 \altaffiliation{keval.physics@yahoo.com}
\author{Ajay Kumar Rai}%
 \email{raiajayk@gmail.com}
\affiliation{%
 Department of Physics, Sardar Vallabhbhai National Institute of Technology, Surat 395007, Gujarat, India 
}%


\author{Takayuki Matsuki}
 \homepage{matsuki@tokyo-kasei.ac.jp}
\affiliation{
 Tokyo Kasai University, 1-18-1 Kaga, Itabashi, Tokyo 173-8602, Japan
}%
\affiliation{
 Theoretical Research Division, Nishina Center, RIKEN, Saitama 351-0198, Japan
}%
%

\date{\today}

\begin{abstract}
Experimentally observed excited strange charmed mesons $D_{s0}^*(2317)^{\pm}$, $D_{s1}(2460)^{\pm}$, $D_{s1}(2536)^{\pm}$, $D_{s2}^*(2573)^{\pm}$, $D_{s1}^*(2700)^{\pm}$, $D_{s1}^*(2860)^{\pm}$, $D_{s3}^*(2860)^{\pm}$, and $D_{sJ}(3040)^{\pm}$ are identified tentatively according to their spin, parity, and masses. Using the heavy quark effective theory in the leading order approximation, we study their strong decays to ground state charmed mesons plus light pseudoscalar mesons. The branching ratios are classified, $(D_{s0}^*(2317), D_{s1}(2460))$ as $(1^3P_0, 1P_1)$ and $(D_{s1}(2536), D_{s2}^*(2573))$ as $(1P_1^{\prime}, 1^3P_2)$, in heavy quark doublets. The $D_{s1}^*(2700)$ as $2^3S_1$ and its spin partner, $2^1S_0$ is still missing in the experiment. Its strong decay behavior is predicted. The assignment of $D_{sJ}^*(2860)$ as four resonance states $D_{s1}^*(2860)$, $D_{s2}^*(2860)$, $D_{s2}^{*\prime}(2860)$, $D_{s3}^*(2860)$ is favored, where for $D_{s3}^*(2860)^{\pm}$ the mass should be expected to higher than 2.86 GeV. The $D_{sJ}(3040)$ is more likely to be 2$P_1$ state. We construct the Regge trajectories of experimentally observed strange charmed mesons in $(M^2, J)$ and $(M^2, n_r)$ planes, which estimate the masses of $1^3D_3$, $1^3F_4$, $1D_2$, $1^1F_3$, $1D_2^{\prime}$, $1F_3^{\prime}$, $3^3S_1$, and $3P_1$ states with fixing the slopes and intercepts of the Regge lines. Moreover, the strong decay rates and the branching ratios of these higher excited states are also examined, which distinguished the dominant decay channels for these higher excited states. Our results could provide some important clues in LHCb, BESIII, KEK-B, and the forthcoming Belle II and $\overline{\mbox{\sffamily P}}${\sffamily ANDA} experiments.
\end{abstract}

\maketitle


\section{Introduction}
\label{sec1}

The latest Review of Particle Physics (RPP) by Particle Data Group (PDG) listed the strange charmed mesons (observed by the various experimental groups) $D_{s0}^*(2317)^{\pm}$, $D_{s1}(2460)^{\pm}$, $D_{s1}(2536)^{\pm}$, $D_{s2}^*(2573)^{\pm}$, $D_{s1}^*(2700)^{\pm}$, $D_{s1}^*(2860)^{\pm}$, $D_{s3}^*(2860)^{\pm}$, and $D_{sJ}(3040)^{\pm}$ with their properties like masses, decay widths, branching ratios, lifetimes, spin, parity etc. \cite{Tanabashi2018-19,Zyla2020}. Since 2003, the Belle, CLEO, $BABAR$, LHC, and BESIII have been producing data on the status of charmed strange mesons. In the upcoming years, we believe that more candidates for excited charmed strange mesons will be reported, with the running experimental facilities LHCb, $BABAR$ and BESIII, and the future experiments Belle II and $\overline{\mbox{\sffamily P}}${\sffamily ANDA}. The comprehensive review articles have given experimental information for charmed mesons \cite{Kato2019,Chen2017,Asner2011}.   
 
More recently, the BESIII Collaboration has studied the $e^+ e^-$ annihilation processes at the center-of-mass energy 4.6 GeV \cite{Ablikim2019}. In the decay mode $e^+ e^- \rightarrow D_{s}^{+}\bar{D}^{(*)0}K^-$, they observed the masses and decay widths of two intermediate $P$-wave charmed strange mesons $D_{s1}(2536)^{-}$ and $D_{s2}^*(2573)^{-}$ with statistical and systematic uncertainties. Their measured spin-parity $2^+$ of $D_{s2}^*(2573)^{-}$ is consistent with PDG \cite{Tanabashi2018-19,Zyla2020}. In 2015, the $BABAR$ Collaboration analyzed the Dalitz plot for the $B$ mesons decays to $D^-D^0K^+$ and $\bar{D}^0D^0K^+$ \cite{Lees2015}. They observed the state $D_{s1}^*(2700)^{+}$ of mass 2699$^{+14}_{-7}$ MeV/c$^2$ and the decay width 127$^{+24}_{-19}$ MeV. The LHCb detector recorded the resonance structure near 2.86 GeV/c$^2$ mass region in the decay mode $B \rightarrow \bar{D}^0K^-\pi^-$ \cite{Aaij2014}. They have found  an admixture of spin-1 and spin-3 resonances, which were separately presented by $D_{s1}^*(2860)^{\pm}$ and $D_{s3}^*(2860)^{\pm}$ states \cite{Tanabashi2018-19,Zyla2020}. The  final state mass spectra analysis of $D^0K^+$ and $D^+K_s^0$ states by LHCb Collaboration have confirmed the existence of $D_{s1}^*(2700)^{+}$ and $D_{sJ}^*(2860)^{+}$ states \cite{Aaij2012}.

\begin{table*}
\caption{\label{tab1}
The experimental results (masses and decay widths) from BESIII(2019) \cite{Ablikim2019}, LHCb(2014) \cite{Aaij2014}, LHCb(2012) \cite{Aaij2012}, $BABAR$(2009) \cite{Aubert2009}, and $BABAR$(2006) \cite{Aubert2006,Aubert2006_2} of strange charmed mesons (in MeV). The notation of strange charmed mesons is taken from PDG \cite{Tanabashi2018-19}.}
\scalebox{1}{
\begin{ruledtabular}
\begin{tabular}{cccccccccc}
Meson & BESIII & LHCb & LHCb & $BABAR$ & $BABAR$ & $BABAR$\\
& (2019) \cite{Ablikim2019} & (2014) \cite{Aaij2014} & (2012) \cite{Aaij2012} & (2009) \cite{Aubert2009} & (2006) \cite{Aubert2006} & (2006) \cite{Aubert2006_2} \\
\noalign{\smallskip}\hline\noalign{\smallskip}
$D_{s0}^*(2317)^{\pm}$ &&&&& 2319.6 $\pm$ 0.2 $\pm$ 1.4 \\
&&&&& $<$3.8\\
&&&&& $0^+$\\
\noalign{\smallskip}\noalign{\smallskip}
$D_{s1}(2460)^{\pm}$ &&&&& 2460.1 $\pm$ 0.2 $\pm$ 0.8  \\
&&&&& $<$3.5\\
&&&&& $1^+$\\
\noalign{\smallskip}\noalign{\smallskip}
$D_{s1}(2536)^{\pm}$ & 2537.7 $\pm$ 0.5 $\pm$ 3.1  \\
& 1.7 $\pm$ 1.2 $\pm$ 0.6\\
& $1^+$\\
\noalign{\smallskip}\noalign{\smallskip}
$D_{s2}^*(2573)^{\pm}$ & 2570.7 ${\pm}$ 2.0 ${\pm}$ 1.7 & 2568.39 ${\pm}$ 0.29 ${\pm}$ 0.26 &&\\
& 17.2 ${\pm}$ 3.6 ${\pm}$ 1.1 & 16.9 ${\pm}$ 0.5 ${\pm}$ 0.6 &&&\\
& $2^+$ & $2^+$\\
\noalign{\smallskip}\noalign{\smallskip}
$D_{s1}^*(2700)^{\pm}$ &&& 2709.2 ${\pm}$ 1.9 ${\pm}$ 4.5 & 2710 ${\pm}$ 2$^{+12}_{-7}$ && 2688 $\pm$ 4 $\pm$ 3\\
&&& 115.8 $\pm$ 7.3 $\pm$ 12.1 & 149 ${\pm}$ 7$^{+39}_{-52}$ && 112 $\pm$ 7 $\pm$ 36 \\
&&& $1^-$ & $1^-$ && Natural\\
\noalign{\smallskip}\noalign{\smallskip}
$D_{s1}^*(2860)^{\pm}$ && 2859 $\pm$ 12 $\pm$ 24 & 2866.1 ${\pm}$ 1.0 ${\pm}$ 6.3 & 2862 $\pm$ 2$^{+5}_{-2}$ && 2856.6 $\pm$ 1.5 $\pm$ 5 \\
&& 159 $\pm$ 23 $\pm$ 77 & 69.9 $\pm$ 3.2 $\pm$ 6.6 & 48 $\pm$ 3 $\pm$ 6&& 48 $\pm$ 7 $\pm$ 10\\
&& $1^-$ & $1^-$ & $1^-$ && Natural\\
\noalign{\smallskip}\noalign{\smallskip}
$D_{s3}^*(2860)^{\pm}$ && 2860.5 $\pm$ 2.6 $\pm$ 6.5\\
&& 53 $\pm$ 7 $\pm$ 7\\
&& $3^-$\\
\noalign{\smallskip}\noalign{\smallskip}
$D_{sJ}(3040)^{\pm}$ &&&& 3044 $\pm$ 8$^{+30}_{-5}$  \\
&&&& 239 $\pm$ 35$^{+46}_{-42}$\\
&&&& Unnatural\\
\end{tabular}
\end{ruledtabular}}
\end{table*}

The $BABAR$ experiment collect the data sample corresponding to an integrated luminosity 470 fb$^{-1}$ at the center-of-mass energy 10.6 GeV of $e^+e^-$ collisions and observed the decays of $D_{s1}^*(2700)^{+}$ and $D_{sJ}^*(2860)^{+}$ to $D^*K$ \cite{Aubert2009}. Their measured branching fraction relative to $DK$ decay modes are,
\begin{equation}
\label{eq:1} 
\frac{{\cal{B}}(D_{s1}^*(2700)^+ \rightarrow D^{*}K)}{{\cal{B}}(D_{s1}^*(2700)^+ \rightarrow DK)} = 0.91 \pm 0.13 \pm 0.12, 
\end{equation} 
\begin{equation}
\label{eq:2} 
\frac{{\cal{B}}(D_{sJ}^*(2860)^+ \rightarrow D^{*}K)}{{\cal{B}}(D_{sJ}^*(2860)^+ \rightarrow DK)} = 1.10 \pm 0.15 \pm 0.19.
\end{equation}

\noindent They also observed a new broad resonant structure in the $D^*K$ invariant mass distribution of unnatural parity ($0^-, 1^+, 2^-, 3^+,...$), having a mass 3044 $\pm$ 8$^{+30}_{-5}$ MeV/c$^2$ and decay width 239 $\pm$ 35$^{+46}_{-42}$ MeV. In Ref. \cite{Aubert2006}, the $BABAR$ Collaboration has studied the final states mass spectra of $D_s^+ \pi^0$, $D_s^+\gamma$ and $D_s^+\pi^+\pi^-$, and observed the first orbital excited states of charmed strange mesons $D_{s0}^*(2317)^{\pm}$ and $D_{s1}(2460)^{\pm}$, using the data 232 fb$^{-1}$ in $e^+e^- \rightarrow \bar{c}c$ decay processes. The $BABAR$ Collaboration \cite{Aubert2006_2} has recorded the data corresponding to 240 fb$^{-1}$ of $e^+e^-$ collisions at center-of-mass energies near 10.6 GeV. They observed $D_S^+$ state decaying to $DK$ at a mass of 2.86 GeV/c$^2$ first time. The decay to two pseudoscalar mesons implies a natural spin-parity ($J^P = 0^+, 1^-, 2^+, ...$) for this state. In the same mass distributions, they also found a broad resonance structure with mass 2.69 GeV/c$^2$ \cite{Aubert2006_2}. Later, such a state was identified as $D_{s1}^*(2700)^+$ \cite{Aubert2009}. 

\begin{table*}
\caption{\label{tab2}
Spectra of strange charmed mesons obtained from different models (in MeV).}
\begin{ruledtabular}
\begin{tabular}{ccccccccccccc}
$\cal{N}$ $^{2S+1}L_J$ & $J^P$ & Ref. \cite{Kher2017} & Ref. \cite{Godfrey2016} & Ref. \cite{Song2015} & Ref. \cite{Li2011} & Ref. \cite{Ebert2010} & Ref. \cite{Yun2010} & Ref. \cite{DiPierro2001} & Ref. \cite{Lahde2000} & Ref. \cite{Cichy2016} \\
\noalign{\smallskip}\hline\noalign{\smallskip}

$1^1S_0$ & $0^-$ & 1953 & 1979 & 1967 & 1969 & 1969 &          & 1965 & 1975 & 1968\\
$1^3S_1$ & $1^-$ & 2112 & 2129 & 2115 & 2107 &  2111 &          & 2113 & 2108 & 2123\\
$2^1S_0$ & $0^-$ & 2642 & 2673 & 2646 & 2640 & 2688 &          & 2700 & 2659 \\
$2^3S_1$ & $1^-$ & 2732 & 2732 & 2704 & 2714 & 2731 &          & 2806 & 2722\\
$3^1S_0$ & $0^-$ & 3219 & 3154 &          &          & 3219 &          &          & 3044\\
$3^3S_1$ & $1^-$ & 3284 & 3193 &          &          & 3242 &          &          & 3087\\

$1^3P_0$ & $0^+$ & 2438 & 2484 & 2463 & 2344 & 2509 & 2478 & 2487 & 2455 & 2390 \\
$1P_1$ & $1^+$ & 2529 & 2549 & 2531 & 2488 & 2536 & 2554 & 2605 & 2502 & 2556\\
$1P_1^{\prime}$ & $1^+$ & 2541 & 2556 & 2532 & 2510 & 2574 & 2516 & 2535 & 2522 & 2617 \\
$1^3P_2$ & $2^+$ & 2569 & 2592 & 2571 & 2559 & 2571 & 2592 & 2581 & 2586 & 2734\\

$2^3P_0$ & $0^+$ & 3022 & 3005 & 2960 & 2830 & 3054 &          & 3067 & 2901\\
$2P_1$ & $1^+$ & 3081 & 3018 & 2979 & 2958 & 3067 &          & 3165 & 2928\\
$2P_1^{\prime}$ & $1^+$ & 3092 & 3038 & 2988 & 2995 & 3154 &          & 3114 & 2942\\
$2^3P_2$ & $2^+$ & 3109 & 3048 & 3004 & 3040 & 3142 &          & 3157 & 2980\\

$3^3P_0$ & $0^+$ & 3541 & 3412 &          &          & 3346 &          &          & 3214 \\
$3P_1$ & $1^+$ & 3587 & 3416 &          &          & 3365 &          &          & 3234\\
$3P_1^{\prime}$ & $1^+$ & 3596 & 3433 &          &          & 3461 &          &          & 3244\\
$3^3P_2$ & $2^+$ & 3609 & 3439 &          &          & 3407 &          &          & 3283\\

$1^3D_1$ & $1^-$ & 2882 & 2899 & 2865 & 2804 & 2913 & 2714 & 2913 & 2838\\
$1D_2$ & $2^-$ & 2853 & 2900 & 2877 & 2788 & 2931 & 2827 & 2953 & 2845 \\
$1D_2^{\prime}$ & $2^-$ & 2872 & 2926 & 2882 & 2849 & 2961 & 2789 & 2900 & 2856\\
$1^3D_3$ & $3^-$ & 2860 & 2917 & 2883 & 2811 & 2971 & 2903 & 2925 & 2857\\

$2^3D_1$ & $1^-$ & 3394 & 3306 & 3244 & 3217 & 3228 &          &          & 3144\\
$2D_2$ & $2^-$ & 3368 & 3298 & 3247 & 3217 & 3259 &          &          & 3167\\
$2D_2^{\prime}$ & $2^-$ & 3384 & 3323 & 3252 & 3260 & 3307 &          &          & 3172\\
$2^3D_3$ & $3^-$ & 3372 & 3311 & 3251 & 3240 & 3335 &          &          & 3157\\

$1^3F_2$ & $2^+$ &        & 3208 & 3159 &          & 3230 & 2894 & 3224\\
$1F_3$ & $3^+$ &        & 3186 & 3151 &          & 3254 & 3046 & 3247\\
$1F_3^{\prime}$ & $3^+$ &        & 3218 & 3157 &          & 3266 & 3008 & 3203\\
$1^3F_4$ & $4^+$ &        & 3190 & 3143 &          & 3300 & 3160 & 3220\\
\end{tabular}
\end{ruledtabular}
\end{table*}

Table \ref{tab1} shows the masses, decay widths and the spin-parity of charmed strange mesons observed by the experimental groups BESIII \cite{Ablikim2019}, LHCb \cite{Aaij2014,Aaij2012}, and  $BABAR$ \cite{Aubert2009,Aubert2006,Aubert2006_2}. Except for $D_{sJ}(3040)^{\pm}$, the $J^P$ ($J$ is the total spin and $P$ is parity) value of all strange charmed mesons is measured experimentally. Because of the limited statistics, the angular distributions for this state have not been studied. For $D_{s0}^*(2317)^{\pm}$ and $D_{s1}(2460)^{\pm}$ the observed masses are far lower than the corresponding results computed using different theoretical approaches. And no one state observed near $D_{s0}^*(2317)^{\pm}$ and $D_{s1}(2460)^{\pm}$ mass, are decaying into $D_s^+\pi^{\pm}$. Also, their charged or neutral states are still not observed experimentally. As it would be expected if the $D_{s0}^*(2317)^{\pm}$ and $D_{s1}(2460)^{\pm}$ are candidates of tetraquark states. Theoretically, the masses of $D_{s0}^*(2317)^{\pm}$ and $D_{s1}(2460)^{\pm}$ states are lies in the range of 2.44 GeV to 2.55 GeV. There have been numbers of theoretical approaches developed to study the heavy-light meson spectroscopy  \cite{Kher2017,Godfrey2016,Song2015,Li2011,Ebert2010,Yun2010,DiPierro2001,Lahde2000,Cichy2016}. 

D. Jiaa and W.-C. Dong derived the Regge-like mass relation of excited singly heavy mesons \cite{Jiaa2019}. In Ref. \cite{Chen2018}, J.-K. Chen constructed the radial and the orbital Regge trajectories for heavy-light mesons by applying the Bohr-Sommerfeld quantization approach, the semi-relativistic approach by V. Kher et al. \cite{Kher2017}, the Godfrey-Isgur (GI) relativized quark model \cite{Godfrey2016}, the GI model with screened potential \cite{Song2015}, the nonrelativistic constituent quark model which proposed by Lakhina and Swanson \cite{Li2011}, the Quantum Chromodynamics (QCD) ­motivated relativistic quark model \cite{Ebert2010}, the mass loaded flux tube model \cite{Yun2010}, the relativistic quark model including the leading order corrections in $1/m$ \cite{DiPierro2001}, the Blankenbecler-Sugar equation in the framework of heavy-light interaction models \cite{Lahde2000}, the lattice QCD \cite{Cichy2016} etc.  

They have computed the masses of the excited state of strange charmed mesons and predict its $J^P$ values are presented in Table \ref{tab2} (the symbol $\cal{N}$ $^{2S+1}L_J$ is used to represent the meson quantum state; where $\cal{N}$, $L$ and $S$ denote the radial, orbital and the intrinsic spin quantum number respectively). The $J^P$ value assignments are important for studying the decay properties of the states. From the mass spectrum analysis of strange charmed mesons we conclude the following points,

\begin{enumerate}[label=\roman*.]
\item The two $1S$ states ($D_s$ and $D^*_s$) and the two $1P$ states ($D_{s1}(2536)$ and $D{_{s2}^*}(2573)$) are theoretically reproduced very well.
\item The calculated masses of the states $D_{s0}^*(2317)^{\pm}$ and $D_{s1}(2460)^{\pm}$ are overestimated to the experimental measurements. 
\item The $D_{s1}^*(2700)^{\pm}$ is measured with $J^P = 1^-$ and can be a good candidate for $2^3S_1$. 
\item The $D_{s1}^*(2860)^{\pm}$ and $D_{s3}^*(2860)^{\pm}$ are observed with the total spin-1 and 3, with negative parity, respectively. So they can be a good candidate for $1D$ states.
\item The $D_{sJ}(3040)^{\pm}$ is observed with unnatural parity. Theoretically, its mass is nearer to 2$P_1$ lower and higher state.
\end{enumerate}

Here, we apply the heavy quark effective theory (HQET) to discuss the quantum number assignments of excited strange charmed mesons reported by BESIII \cite{Ablikim2019}, LHCb \cite{Aaij2014,Aaij2012}, and $BABAR$ \cite{Aubert2009,Aubert2006,Aubert2006_2}. HQET was originally proposed to study the two-body strong interaction of heavy-light mesons with the emission of light pseudoscalar mesons $(\pi, \eta,$ and $K)$ \cite{FalkandLuke1992,Falk1992,Wiseandall1992,Casalbuoni1992,Ebert1995}. In HQET, the spin-flavor symmetry of the heavy quark reduced the number of unknown parameters and the heavy quark four-vector $\nu_{\mu}$ in the superfield equations conserved the strong interaction processes at the infinite heavy quark mass limit, which makes it easier to study the properties of heavy-light systems. In the past two decades, many theoretical groups have successfully explained the properties of heavy-light mesons in the framework of HQET \cite{Colangelo2006_1,Colangelo2006_2,Colangelo2008,Colangelo2010,Colangelo2012,Wang2011,Wang2012,Wang2013,Batra2015,Gupta2018,Gandhi2019arxiv}.  P. Colangelo, F. De Fazio, and S. Nicotri \cite{Colangelo2006_2} analyzed $D_{sJ}^*(2860)$ meson (discovered by $BABAR$ experiment \cite{Aubert2006_2}) in the effective Lagrangian approach based on heavy quark spin-flavor and light quark chiral symmetry. They identify $D_{sJ}^*(2860)$ with spin-parity $3^-$. In Ref. \cite{Colangelo2012}, the same group computed the strong decay rates and the ratios of branching fractions of the experimentally observed open charm and open bottom mesons. They continue to propose for $D_{sJ}^*(2860)$ as $1^3D_3$, and predict the mass of its spin partner $D_{s2}^{\prime*}$, from $M_{D_{s2}^{\prime*}} = M_{D_{s3}} - (M_{D_3} - M_{D_{2}^{\prime*}})$. Using the assumption that $D_{sJ}^*(2860)$ and $D_3^*(2750)$ have the same quantum numbers and differ for the strangeness. In Ref. \cite{Wang2015}, they take $D_{s1}^*(2860)$ as $1^3D_1$ and $D_{s3}^*(2860)$ as $1^3D_3$, and studied their strong decays with heavy meson effective theory including the chiral symmetry-breaking corrections. They also reproduce the experimental value of the ratio 1.10 $\pm$ 0.15 $\pm$ 0.19 \cite{Aubert2009} with suitable coupling constants and assign the $D_{sJ}^*(2860)$ as $D_{s3}^*(2860)$, the chiral symmetry-breaking corrections are large.

In Ref. \cite{Wang2015}, Wang says if the chiral symmetry-breaking corrections are small, then such large ratio 1.10 $\pm$ 0.15 $\pm$ 0.19 requires that $D_{sJ}^*(2860)$ consists of at least four resonance states $D_{s1}^*(2860)$, $D_{s2}^*(2860)$, $D_{s2}^{*\prime}(2860)$, $D_{s3}^*(2860)$. This is the motivation of our present work. In this paper, the branching ratio of the decay widths predicting the quantum states of $D_s$ mesons $D_{s0}^*(2317)^{\pm}$, $D_{s1}(2460)^{\pm}$, $D_{s1}(2536)^{\pm}$, $D_{s2}^*(2573)^{\pm}$, and $D_{s1}^*(2700)^{\pm}$. The mixing angle calculation identifies the $D_{sJ}(3040)$ state as an admixture of $2P_1 - 2P_1^{\prime}$, but $2P_1$ is favored. This makes it possible to construct their Regge trajectories in $(M^2, J)$ and $(M^2, n_r)$ planes with $D_{s1}^*(2860)^{\pm}$ as $1^3D_1$ and $D_{s3}^*(2860)^{\pm}$ as $1^3D_3$. We fixed the slopes and intercepts of the Regge lines and estimated the masses of higher excited strange charmed mesons $1^3D_3$, $1^3F_4$, $1D_2$, $1^1F_3$, $1D_2^{\prime}$, $1F_3^{\prime}$, $3^3S_1$, and $3P_1$. Moreover, their ratio among the decay rates is also predicted. These predictions can provide some crucial information for upcoming experimental studies.

This paper is divided into the following sections. Section \ref{sec2} describes the heavy quark effective theory formalism used for the strong decays. Section \ref{sec3} presents the numerical analysis of the strong decay rates, where we attempt to identify the spin-parity of experimentally known excited strange charmed mesons. In section \ref{sec4} we plot the Regge trajectories in $(M^2, J)$ and $(M^2, n_r)$planes using the masses from PDG \cite{Tanabashi2018-19,Zyla2020}. Further, we examined the ratios of the strong decay rates of the states ($1^3D_3$, $1^3F_4$, $1^1D_2$, $1^1F_3$, $1^3D_2$, $1^3F_3$, $3^3S_1$, and $3P_1$) lying on the Regge lines. At last, in section \ref{sec5} we  summarized this work.

\section{Theoretical framework}
\label{sec2}

In the heavy quark effective theory (HQET) the masses of the light $u$, $d$, and $s$ quarks, say $m_q$, are small compared to the scale of non-perturbative strong dynamics, i.e. $m_q \rightarrow 0$ limit of QCD. In this limit, QCD has chiral symmetry, which can be used to predict some properties of hadrons containing these light quarks. And, the heavy $c$ and $b$ quark masses are large compared to the scale of non-perturbative strong dynamics, i.e. $m_Q \rightarrow \infty$ limit of QCD. In this heavy quark mass limit QCD has spin-flavor symmetry, which has important implications for the determination of hadrons properties containing a single heavy quark. 

In the $m_Q \rightarrow \infty$ limit, the heavy quark spin $\vec{s}_Q$ is conserved. Therefore, the spin of the light degrees of freedom is $\vec{s}_l =  \vec{s}_{\bar{q}} + \vec{L}$, where $\vec{s}_{\bar{q}}$ and $\vec{L}$ are the spin and the orbital angular momentum of the light antiquark, respectively \cite{Neubert1994}. The heavy-light mesons are classified in doublets in $m_Q \rightarrow \infty$ limit corresponding to $L = 0, 1, 2$, and $3$ for $S$, $P$, $D$, and $F$-waves, respectively. For the ground state, $L = 0$ ($S$-wave) gives $\vec{s}_l^P = {\frac{1}{2}}^-$, has only one doublet containing two states $J^P_{s_l} = (0^-, 1^-)_{{\frac{1}{2}}^-}$ represented by $(P, P^*)$. For $L = 1$ ($P$-wave) has two doublets $\vec{s}_l^P = {\frac{1}{2}}^+$ and $\vec{s}_l^P = {\frac{3}{2}}^+$, have the spin-parity $J^P_{s_l} = (0^+, 1^+)_{{\frac{1}{2}}^+}$ and $J^P_{s_l} = (1^+, 2^+)_{{\frac{3}{2}}^+}$ represented by $(P^*_0, P^{\prime}_1)$ and $(P_1, P_2^*)$, respectively. In the same way for $L = 2$ ($D$-wave) $\vec{s}_l^P = {\frac{3}{2}}^-$ and $\vec{s}_l^P = {\frac{5}{2}}^-$, having $J^P_{s_l} = (1^-, 2^-)_{{\frac{3}{2}}^-}$ and $J^P_{s_l} = (2^-, 3^-)_{{\frac{5}{2}}^-}$ are represented by $(P^*_1, P_2)$ and $(P^{\prime}_2, P^*_3)$, respectively. And, for the $F$-wave ($L = 3$), two doublets $\vec{s}_l^P = {\frac{5}{2}}^+$ and $\vec{s}_l^P = {\frac{7}{2}}^+$, having $J^P_{s_l} = (2^+, 3^+)_{{\frac{5}{2}}^+}$ and $J^P_{s_l} = (3^+, 4^+)_{{\frac{7}{2}}^+}$ are represented by $(P^{\prime*}_2, P_3)$ and $(P^{\prime}_3, P^*_4)$, respectively. Hence, each doublet contains two states (or two spin partners) with total spin $J = s_l \pm \frac{1}{2}$ and parity $P = (-1)^{L+1}$ and can be described by the superfields $H_a$, $S_a$, $T_a$, $X_a$, $Y_a$, $Z_a$, and $R_a$, given by \cite{Wang2013}

\begin{equation}
\label{eq:3} 
H_a = \frac{1 + {\rlap{v}/}}{2} [P^*_{a\mu}\gamma^{\mu} - P_a\gamma_5],
\end{equation}
\begin{equation}
\label{eq:4} 
S_a = \frac{1 + {\rlap{v}/}}{2} [P^{\mu}_{1a}\gamma_{\mu}\gamma_5 - P^*_{0a}],
\end{equation}
\begin{equation}
\label{eq:5} 
T_a^{\mu} = \frac{1 + {\rlap{v}/}}{2} \Bigg\{P_{2a}^{*\mu\nu} \gamma_{\nu} - P_{1a\nu} \sqrt{\frac{3}{2}} \gamma_5 \bigg[g^{\mu\nu} - \frac{\gamma^{\nu}(\gamma^{\mu} - v^{\mu})}{3} \bigg]\Bigg\},
\end{equation}
\begin{equation}
\label{eq:6} 
X_a^{\mu} = \frac{1 + {\rlap{v}/}}{2} \Bigg\{P_{2a}^{\mu\nu} \gamma_{5} \gamma_{\nu} - P_{1a\nu}^{*} \sqrt{\frac{3}{2}} \bigg[g^{\mu\nu} - \frac{\gamma^{\nu}(\gamma^{\mu} + v^{\mu})}{3} \bigg]\Bigg\},
\end{equation}
\begin{equation}
\begin{aligned}
\label{eq:7} 
Y_a^{\mu\nu} = & \frac{1 + {\rlap{v}/}}{2} \Bigg\{P_{3a}^{*\mu\nu\sigma} \gamma_{\sigma} - P_{2a}^{\alpha\beta} \sqrt{\frac{5}{3}} \gamma_5\\
& \bigg[g^{\mu}_{\alpha} g^{\nu}_{\beta} - \frac{g^{\nu}_{\beta}\gamma_{\alpha}(\gamma^{\mu} - v^{\mu})}{5} - \frac{g_{\alpha}^{\mu}\gamma_{\beta}(\gamma^{\mu} - v^{\nu})}{5} \bigg]\Bigg\},
 \end{aligned}
\end{equation}
\begin{equation}
\begin{aligned}
\label{eq:8} 
Z_a^{\mu\nu} = & \frac{1 + {\rlap{v}/}}{2} \Bigg\{P_{3a}^{\mu\nu\sigma} \gamma_5 \gamma_{\sigma} - P_{2a}^{*\alpha\beta} \sqrt{\frac{5}{3}}\\
& \bigg[g^{\mu}_{\alpha} g^{\nu}_{\beta} - \frac{g^{\nu}_{\beta}\gamma_{\alpha}(\gamma^{\mu} + v^{\mu})}{5} - \frac{g_{\alpha}^{\mu}\gamma_{\beta}(\gamma^{\mu} + v^{\nu})}{5} \bigg]\Bigg\},
\end{aligned}
\end{equation}
\begin{equation}
\begin{aligned}
\label{eq:9} 
R_a^{\mu\nu\rho} = & \frac{1 + {\rlap{v}/}}{2} \Bigg\{P_{4a}^{*\mu\nu\rho\sigma} \gamma_5 \gamma_{\sigma} - P_{3a}^{\alpha\beta\tau} \sqrt{\frac{7}{4}} \\ 
& \bigg[g^{\mu}_{\alpha} g^{\nu}_{\beta} g^{\rho}_{\tau} - \frac{g^{\nu}_{\beta}g^{\rho}_{\tau}\gamma_{\alpha}(\gamma^{\mu} - v^{\mu})}{7} \\
& - \frac{g_{\alpha}^{\mu}g^{\rho}_{\tau}\gamma_{\beta}(\gamma^{\nu} - v^{\nu})}{7} - \frac{g^{\mu}_{\alpha}g^{\nu}_{\beta}\gamma_{\tau}(\gamma^{\rho} -v^{\rho})}{7} \bigg]\Bigg\}.
\end{aligned}
\end{equation}

\noindent where $a$ ($= u, d$ or $s$) is the $SU(3)$ light quark flavor representation and $\nu$ gives the meson four velocity and is conserved in strong interactions. The heavy meson field operators $P$ and $P^*$ (see Eqs. (\ref{eq:3}) to (\ref{eq:9})) contain a factor $\sqrt{m_Q}$ having a mass dimension $\frac{3}{2}$, which annihilate the mesons with four-velocity $\nu$. The field $H_a$ is for $S$-wave doublet $J^P_{s_l} = (0^-, 1^-)_{{\frac{1}{2}}^-}$; fields $S_a$ and $T_a$ are for $P$-wave doublets $J^P_{s_l} = (0^+, 1^+)_{{\frac{1}{2}}^+}$ and $J^P_{s_l} = (1^+, 2^+)_{{\frac{3}{2}}^+}$, respectively. The fields $X_a$ and $Y_a$ are for $D$-wave doublets $J^P_{s_l} = (1^-, 2^-)_{{\frac{3}{2}}^-}$ and $J^P_{s_l} = (2^-, 3^-)_{{\frac{5}{2}}^-}$, respectively; and the fields $Z_a$ and $R_a$ represents the $F$-wave doublets $J^P_{s_l} = (2^+, 3^+)_{{\frac{5}{2}}^+}$ and $J^P_{s_l} = (3^+, 4^+)_{{\frac{7}{2}}^+}$, respectively. The $SU(3)$ chiral symmetry of massless three-light flavor QCD is spontaneously broken in this non-perturbative strong interaction dynamics. The eight broken $SU(3)$ generators transform the composite field along symmetry direction. Fluctuations in the field space along these eight directions are eight light Goldstone bosons. That is described by the fields $\xi = e^{\frac{i\cal{M}}{f_{\pi}}}$, where $f_{\pi} = 130.2$ MeV is the pion decay constant and $\cal{M}$ is written in the form of 3$\times$3 matrix as  

\begin{equation}
\label{eq:10} 
 \cal{M} = 
\begin{pmatrix}
\frac{1}{\sqrt{2}}\pi^0 + \frac{1}{\sqrt{6}}\eta & \pi^+ & K^+ \\
\pi^- & -\frac{1}{\sqrt{2}}\pi^0 + \frac{1}{\sqrt{6}}\eta & K^0 \\
K^- & \bar{K}^0 & -\sqrt{\frac{2}{3}}\eta \\
\end{pmatrix}.
\end{equation}

\noindent In the leading order approximation, the effective heavy meson chiral Lagrangians $\cal{L}$$_H$, $\cal{L}$$_S$, $\cal{L}$$_T$, $\cal{L}$$_X$ $\cal{L}$$_Y$, $\cal{L}$$_Z$, and $\cal{L}$$_R$ describe the two-body strong interactions by an exchange of light pseudoscalar mesons are taken from \cite{Wang2013,Wiseandall1992},

\begin{equation}
\label{eq:11} 
{\cal{L}}_H = g_HTr[\bar{H}_aH_b\gamma_{\mu}\gamma_5{\cal{A}}^{\mu}_{ba}],
\end{equation}
\begin{equation}
\label{eq:12} 
{\cal{L}}_S = g_STr[\bar{H}_aS_b\gamma_{\mu}\gamma_5{\cal{A}}^{\mu}_{ba}] + H.C.,
\end{equation}
\begin{equation}
\label{eq:13} 
{\cal{L}}_T = \frac{g_T}{{\Lambda}}Tr[\bar{H}_aT^{\mu}_b(iD_{\mu} {\not\! {\cal A}} + i {\not\! {\cal D}} {\cal{A}}^{\mu})_{ba}\gamma_5] + H.C., 
\end{equation}
\begin{equation}
\label{eq:14} 
{\cal{L}}_X = \frac{g_X}{{\Lambda}}Tr[\bar{H}_aX^{\mu}_b(iD_{\mu} {\not\! {\cal A}} + i {\not\! {\cal D}} {\cal{A}}^{\mu})_{ba}\gamma_5] + H.C., 
\end{equation}
\begin{equation}
\begin{aligned}
\label{eq:15} 
{\cal{L}}_Y = & \frac{1}{{\Lambda}^2}Tr[\bar{H}_aY^{\mu\nu}_b [k_1^Y\{D_{\mu}, D_{\nu}\} {\cal{A}}_{\lambda}\\  
& + k_2^Y(D_{\mu}D_{\lambda}{\cal{A}}_{\nu} + D_{\nu}D_{\lambda}{\cal{A}}_{\mu})]_{ba}\gamma^{\lambda}\gamma_5] + H.C., 
\end{aligned}
\end{equation}
\begin{equation}
\begin{aligned}
\label{eq:16} 
{\cal{L}}_Z = & \frac{1}{{\Lambda}^2}Tr[\bar{H}_aZ^{\mu\nu}_b [k_1^Z\{D_{\mu}, D_{\nu}\} {\cal{A}}_{\lambda}\\
& + k_2^Z(D_{\mu}D_{\lambda}{\cal{A}}_{\nu} + D_{\nu}D_{\lambda}{\cal{A}}_{\mu})]_{ba}\gamma^{\lambda}\gamma_5] + H.C., 
\end{aligned}
\end{equation}
\begin{equation}
\begin{aligned}
\label{eq:17} 
{\cal{L}}_R = & \frac{1}{{\Lambda}^3}Tr[\bar{H}_aR^{\mu\nu\rho}_b [k_1^R\{D_{\mu}, D_{\nu}, D_{\rho}\} {\cal{A}}_{\lambda}\\
& + k_2^R (\{D_{\mu}, D_{\rho}\}D_{\lambda}{\cal{A}}_{\nu} \\
& + \{D_{\nu}, D_{\rho}\}D_{\lambda} {\cal{A}}_{\mu} \{D_{\mu}, D_{\nu}\} D_{\lambda} {\cal{A}}_{\rho})]_{ba} \gamma^{\lambda} \gamma_5 \\
& + H.C.,
\end{aligned} 
\end{equation}
\noindent where vector and axial-vector operators,
\begin{equation}
\begin{aligned}
\label{eq:18} 
{\cal{V}}_{{\mu}ba} = \frac{1}{2} (\xi^{\dag} \partial_{\mu} \xi + \xi \partial_{\mu} \xi^{\dag})_{ba},
\end{aligned} 
\end{equation}
\begin{equation}
\begin{aligned}
\label{eq:19} 
{\cal{A}}_{{\mu}ba} = \frac{i}{2} (\xi^{\dag} \partial_{\mu} \xi - \xi \partial_{\mu} \xi^{\dag})_{ba};
\end{aligned} 
\end{equation}

\noindent where ${\cal{V}}_{{\mu}ba} = \frac{1}{2} (\xi^{\dag} \partial_{\mu} \xi + \xi \partial_{\mu} \xi^{\dag})_{ba}$ and ${\cal{A}}_{{\mu}ba} = \frac{i}{2} (\xi^{\dag} \partial_{\mu} \xi - \xi \partial_{\mu} \xi^{\dag})_{ba}$ are the vector and axial-vector currents. The operator, {\small{$D_{{\mu}ba} = -\delta_{ba} \partial_{\mu} + {\cal{V}}_{{\mu}ba}$}}. Here, {\small{$\{D_{\mu}, D_{\nu}\} = D_{\mu}D_{\nu}+D_{\nu}D_{\mu}$}} and {\small{$\{D_{\mu}, D_{\nu}, D_{\rho}\} = D_{\mu}D_{\nu}D_{\rho} + D_{\mu}D_{\rho}D_{\nu} + D_{\nu}D_{\mu}D_{\rho} + D_{\nu}D_{\rho}D_{\mu} + D_{\rho}D_{\mu}D_{\nu} + D_{\rho}D_{\nu}D_{\mu}$}}. $\Lambda$ is the chiral symmetry breaking scale and is fixed to 1 GeV. The mass parameters $\delta{m}_S = m_S - m_H$, $\delta{m}_T = m_T - m_H$, $\delta{m}_X = m_X - m_H$, $\delta{m}_Y = m_Y - m_H$, $\delta{m}_Z = m_Z - m_H$, and $\delta{m}_R = m_R - m_H$ represent the mass splittings between the higher and the lower mass doublets described by the field $H_a$ (see Eq. (\ref{eq:3})). The strong effective coupling constants $g_H$, $g_S$, $g_T$, $g_X$, $g_Y = k_1^Y + k_2^Y$, $g_Z = k_1^Z + k_2^Z$, and $g_R = k_1^R + k_2^R$ can be fitted to the experimental data. The $g_H$ controls the $S$-wave decays, $g_S$ and $g_T$ are govern the $P$-wave decays, $g_X$ and $g_Y$ describe the $D$-wave decays and, $g_Z$ and $g_R$ are responsible for the $F$-wave decays. The heavy meson chiral Lagrangians with subscript notations $H, S, T, X, Y, Z$, and $R$ indicate the interaction between the super-field $H$ and super-fields $H, S, T, X, Y, Z$, and $R$, respectively. The strong decays of excited charmed mesons into ground state negative parity charmed mesons with the emission of light vector mesons ($\rho$, $\omega$, $K^*$, and $\phi$) are found in Ref. \cite{Campanella2018}. Such a chiral Lagrangians are determined the expressions of strong decays of heavy-light mesons into the ground state charged and neutral charmed mesons along with the light pseudoscalar mesons ($\pi$, $\eta$, and $K$),

\begin{equation}
\label{eq:20}
\Gamma = \frac{1}{2J+1} \sum \frac{\vec{P}_{\cal{P}}}{8\pi P_a^2} |{\cal{M}}|^2
\end{equation}

For the decay mode $P_a \rightarrow P_b + {\cal{P}}$, ${\cal{P}}$ is the emitting light pseudoscalar mesons ($\pi, \eta,$ and $K$) in the strong decay processes. Its final momentum is,

\begin{equation}
\label{eq:21}
|{P}_{\cal{P}}| = \frac{\sqrt{\big(M_{P_a}^2-(M_{P_b} + m_{\cal{P}})^2\big)\big(M_{P_a}^2-(M_{P_b} - m_{\cal{P}})^2\big)}}{2M_{p_a}},
\end{equation}

\noindent where $M_{P_a}$, $M_{P_b}$ and $m_{\cal{P}}$ are their respective masses; $a$ and $b$ denote the initial and final state heavy-light mesons, respectively; $J$ represents the total angular momentum of the initial state of mesons; and $\sum$ gives the summation of all the polarization vectors of the total angular momentum $j =$ 1, 2, 3, or 4; and ${\cal{M}}$ denotes the scattering amplitudes \cite{Wang2013}. The masses of the light pseudoscalar mesons and the ground state charmed mesons are taken from PDG \cite{Tanabashi2018-19,Zyla2020}: $M_{\pi^{\pm}} = 139.57$ MeV, $M_{\pi^0} = 134.98$ MeV, $M_{K^{\pm}} = 493.68$ MeV, $M_{K^{0}} = 497.61$ MeV, $M_{\eta} = 547.86$ MeV, $M_{D^{\pm}} = 1869.65$ MeV, $M_{D^{0}} = 1864.84$ MeV, $M_{D^{*\pm}} = 2010.26$ MeV, $M_{D^{*0}} = 2006.85$ MeV, $M_{D_s^{\pm}} = 1969.00$ MeV, $M_{D_s^{*\pm}} = 2112.20$ MeV.

\begin{table*}
\caption{\label{tab3}
The strong decay widths of the experimentally observed strange charmed mesons with possible quantum state assignments (in MeV), where the first decay channel of each state have been observed experimentally (shown in Table \ref{tab1}).}
\begin{ruledtabular}
\begin{tabular}{ccccccccccccc}
Meson & $\cal{N}$ $^{2S+1}L_J$ & Decay & BESIII & LHCb & LHCb & $BABAR$ & $BABAR$ & $BABAR$ \\
& & mode & (2019) \cite{Ablikim2019} & (2014) \cite{Aaij2014} & (2012) \cite{Aaij2012} & (2009) \cite{Aubert2009} & (2006) \cite{Aubert2006} & (2006) \cite{Aubert2006_2} \\
\noalign{\smallskip}\hline\noalign{\smallskip}
$D_{s0}^*(2317)^{\pm}$ &1$^3P_0$&$D^{0}K^{\pm}$ &&&&& $-$\\
&& $D^{\pm}K^0$ &&&&& $-$\\
&& $D_s^{\pm}\pi^0$ &&&&& 128.19$g_S^2$ \\
&&&&&&& $\times$ 10$^{-4}$\\
&& $D_s^{\pm}\eta$ &&&&& $-$ \\
\noalign{\smallskip}\hline\noalign{\smallskip} 
$D_{s1}(2460)^{\pm}$ & 1$P_1$ & $D^{*0}K^{\pm}$ &&&&& $-$\\
&& $D^{*\pm}K^0$ &&&&& $-$ \\
&& $D_s^{*\pm}\pi^0$ &&&&& 128.36$g_S^2$\\
&&&&&&& $\times$ 10$^{-4}$\\
&& $D_s^{*\pm}\eta$ &&&&& $-$\\
\noalign{\smallskip}\hline\noalign{\smallskip} 
$D_{s1}(2536)^{\pm}$ & 1$P_1^{\prime}$ & $D^{*0}K^{\pm}$ & 1.56$g_T^2$ \\
&& $D^{*\pm}K^0$ & 0.91$g_T^2$ \\
&& $D_s^{*\pm}\pi^0$ & 35.90$g_T^2$ \\
&&& $\times$ 10$^{-4}$\\
&& $D_s^{*\pm}\eta$ & $-$\\
\noalign{\smallskip}\hline\noalign{\smallskip} 
$D_{s2}^*(2573)^{\pm}$ & 1$^3P_2$ & $D^{*0}K^{\pm}$ & 4.78$g_T^2$ & 4.38$g_T^2$\\ 
&& $D^{*\pm}K^0$ & 3.65$g_T^2$ & 3.31$g_T^2$\\
&& $D_s^{*\pm}\pi^0$ & 62.36$g_T^2$ & 60.84$g_T^2$ \\
&&& $\times$ 10$^{-4}$ & $\times$ 10$^{-4}$\\
&& $D_s^{*\pm}\eta$ & $-$ & $-$\\
&& $D^{0}K^{\pm}$ & 54.99$g_T^2$ & 53.40$g_T^2$\\
&& $D^{\pm}K^0$ & 51.92$g_T^2$ & 50.38$g_T^2$\\
&& $D_s^{\pm}\pi^0$ & 71.18$g_T^2$ & 69.95$g_T^2$\\
&&& $\times$ 10$^{-4}$ & $\times$ 10$^{-4}$\\
&& $D_s^{\pm}\eta$ & 1.26$g_T^2$ & 1.13$g_T^2$\\
\noalign{\smallskip}\hline\noalign{\smallskip} 
$D_{s1}^*(2700)^{\pm}$ & 2$^3S_1$ & $D^{*0}K^{\pm}$ &&& 374.92$g_H^2$ & 377.20$g_H^2$ && 316.38$g_H^2$ \\ 
&& $D^{*\pm}K^0$ &&& 356.82$g_H^2$ & 359.08$g_H^2$ && 299.26$g_H^2$ \\
&& $D_s^{*\pm}\pi^0$ &&& 337.68$g_H^2$ & 338.88$g_H^2$ && 306.67$g_H^2$\\
&&&&& $\times$ 10$^{-4}$ & $\times$ 10$^{-4}$ && $\times$ 10$^{-4}$ \\
&& $D_s^{*\pm}\eta$ &&& 29.99$g_H^2$ & 30.74$g_H^2$ && 12.73$g_H^2$ \\
&& $D^{0}K^{\pm}$ &&& 409.27$g_H^2$ & 410.77$g_H^2$ && 370.26$g_H^2$ \\
&& $D^{\pm}K^0$ &&& 401.28$g_H^2$ & 402.77$g_H^2$ && 362.53$g_H^2$ \\
&& $D_s^{\pm}\pi^0$ &&& 281.80$g_H^2$ & 282.56$g_H^2$ && 262.10$g_H^2$ \\
&&&&& $\times$ 10$^{-4}$ & $\times$ 10$^{-4}$ && $\times$ 10$^{-4}$ \\
&& $D_s^{\pm}\eta$ &&& 117.88$g_H^2$ & 118.64$g_H^2$ && 98.21$g_H^2$ \\
\noalign{\smallskip}\hline\noalign{\smallskip} 
$D_{s1}^*(2860)^{\pm}$ & 1$^3D_1$ & $D^{*0}K^{\pm}$ && 351.04$g_X^2$ & 367.30$g_X^2$ & 357.86$g_X^2$ &&   345.43$g_X^2$  \\ 
&& $D^{*\pm}K^0$ && 340.20$g_X^2$ & 356.19$g_X^2$ & 346.90$g_X^2$ && 334.69$g_X^2$ \\
&& $D_s^{*\pm}\pi^0$ && 170.67$g_X^2$ & 177.70$g_X^2$ & 173.62$g_X^2$ && 168.24$g_X^2$ \\
&&&& $\times$ 10$^{-4}$ & $\times$ 10$^{-4}$ & $\times$ 10$^{-4}$ && $\times$ 10$^{-4}$ \\
&& $D_s^{*\pm}\eta$ && 85.35$g_X^2$ &91.58$g_X^2$ & 87.95$g_X^2$ && 83.22$g_X^2$ \\
&& $D^{0}K^{\pm}$ && 1447.94$g_X^2$ & 1498.51$g_X^2$ & 1469.17$g_X^2$ && 1430.4$g_X^2$  \\
&& $D^{\pm}K^0$ && 1418.83$g_X^2$ & 1468.69$g_X^2$ & 1439.80$g_X^2$ && 1401.51$g_X^2$ \\
&& $D_s^{\pm}\pi^0$ && 669.92$g_X^2$ & 692.17$g_X^2$ & 679.25$g_X^2$ && 662.20$g_X^2$\\
&&&& $\times$ 10$^{-4}$ & $\times$ 10$^{-4}$ & $\times$ 10$^{-4}$ && $\times$ 10$^{-4}$ \\
&& $D_s^{\pm}\eta$ && 504.50$g_X^2$ & 527.93$g_X^2$ & 514.32$g_X^2$ && 496.42$g_X^2$ \\
\noalign{\smallskip}
\end{tabular}
\end{ruledtabular}\\
{continued...}
\end{table*}

\begin{table*}
\addtocounter{table}{-1}
\caption{\label{tab3}
The strong decay widths of the experimentally observed strange charmed mesons with possible quantum state assignments (in MeV), where the first decay channel of each state have been observed experimentally (shown in Table \ref{tab1}).}
\begin{ruledtabular}
\begin{tabular}{ccccccccccccc}
Meson & $\cal{N}$ $^{2S+1}L_J$ & Decay & BESIII & LHCb & LHCb & $BABAR$ & $BABAR$ & $BABAR$ \\
& & mode & (2019) \cite{Ablikim2019} & (2014) \cite{Aaij2014} & (2012) \cite{Aaij2012} & (2009) \cite{Aubert2009} & (2006) \cite{Aubert2006} & (2006) \cite{Aubert2006_2} \\
\noalign{\smallskip}\hline\noalign{\smallskip} 
$D_{s3}^*(2860)^{\pm}$ &1$^3D_3$& $D^{*0}K^{\pm}$ && 49.63$g_Y^2$ \\
&& $D^{*\pm}K^0$ && 46.52$g_Y^2$\\
&& $D_s^{*\pm}\pi^0$ && 46.21$g_Y^2$ \\
&&&& $\times$ 10$^{-4}$\\
&& $D_s^{*\pm}\eta$ && 4.52$g_Y^2$\\
&& $D^{0}K^{\pm}$ && 127.57$g_Y^2$\\
&& $D^{\pm}K^0$ && 123.29$g_Y^2$\\
&& $D_s^{\pm}\pi^0$ && 92.86$g_Y^2$\\
&&&& $\times$ 10$^{-4}$\\
&& $D_s^{\pm}\eta$ && 24.47$g_Y^2$\\
\noalign{\smallskip}\hline\noalign{\smallskip}
$D_{sJ}^*(3040)^{\pm}$ &2$P_1$& $D^{*0}K^{\pm}$ &&&& 3779.96$g_S^2$ \\
&& $D^{*\pm}K^0$ &&&& 3750.19$g_S^2$\\
&& $D_s^{*\pm}\pi^0$ &&&& 1595.61$g_S^2$ \\
&&&&&& $\times$ 10$^{-4}$ \\
&& $D_s^{*\pm}\eta$ &&&& 1938.16$g_S^2$\\
\noalign{\smallskip}\hline\noalign{\smallskip} 
$D_{sJ}^*(3040)^{\pm}$ &2$P_1^{\prime}$& $D^{*0}K^{\pm}$ &&&& 1999.37$g_T^2$ \\
&& $D^{*\pm}K^0$ &&&& 1943.32$g_T^2$ \\
&& $D_s^{*\pm}\pi^0$ &&&& 1258.69$g_T^2$\\
&&&&&& $\times$ 10$^{-4}$ \\
&& $D_s^{*\pm}\eta$ &&&& 596.28$g_T^2$\\
\noalign{\smallskip}
\end{tabular}
\end{ruledtabular}
\end{table*}

For $D_{s3}^*(2860)$ as $1^3D_3$, the two-body strong decays $D_{s3}^*(2860) \rightarrow DK, D^*K$; having the $K$ mesons three momenta $\vec{P}_K =$ 589 and 710 MeV, respectively. The decay widths    

\begin{equation}
\label{eq:22}
\Gamma(D_{s3}^*(2860) \rightarrow DK, D^*K) \propto \vec{P}_{K}^7,
\end{equation}

\noindent where $\vec{P}_{K}^7 = $ 9.1 $\times$ 10$^{19}$  and 2.5 $\times$ 10$^{19}$ MeV$^7$ in the decays to the final states $DK$ and $D^*K$, respectively. A small difference in $\vec{P}_K$ can lead to a large difference in $\vec{P}_{K}^7$, so we have to take into account the heavy quark symmetry-breaking corrections and chiral symmetry-breaking corrections so as to make robust predictions \cite{Wang2013}. The higher-order corrections for spin and flavor violation of the order ${\cal{O}}(\frac{1}{m_Q})$ are not taking into consideration to avoid introducing new unknown coupling constants. We expect that the corrections would not be larger than (or as large as) the leading order contributions \cite{Wang2015}. At the hadronic level, the $\frac{1}{m_Q}$ corrections can be crudely estimated to be of the order $\frac{\vec{P}_{K}}{M_{D_s}} \approx$ 0.1-0.3.

\section{Numerical analysis}
\label{sec3}

The numerical values of the strong decay rates of the excited strange charmed mesons $D_{s0}^*(2317)^{\pm}$, $D_{s1}(2460)^{\pm}$, $D_{s1}(2536)^{\pm}$, $D_{s2}^*(2573)^{\pm}$, $D_{s1}^*(2700)^{\pm}$, $D_{s1}^*(2860)^{\pm}$, $~~~~$ $D_{s3}^*(2860)^{\pm}$, and $D_{sJ}^*(3040)^{\pm}$ observed by the  BESIII \cite{Ablikim2019},  LHCb \cite{Aaij2014,Aaij2012}, and $BABAR$ \cite{Aubert2009,Aubert2006} Collaborations are presented in Table \ref{tab3}, where the strong decay widths are retained in the form of a square of the couplings $g_H$, $g_S$, $g_T$, $g_X$, $g_Y$, $g_Z$, and $g_R$. Without enough experimental information, it is impossible to determine their values with heavy quark effective theory. The running experimental facilities LHCb, BESIII, KEK-B, and the future project $\overline{\mbox{\sffamily P}}${\sffamily ANDA} \cite{Singh(all),Barucca2019} are expected to fit such strong couplings and can be studied the heavy meson interactions in the near future. Theoretically, the Refs. \cite{Fajfer2006,Stewart1998} have used the combined approach of heavy meson chiral perturbation theory with heavy quark effective theory and extracted the coupling $g_H$ for the decays of $D^{*0}$, $D^{*+}$, $D^*_s$ to $D\pi$ mode. The Refs. \cite{Colangelo1995,Casalbuoni1997,Wang2006,Wang2007} are mainly focused on the coupling constants $g_H$, $g_S$, and $g_T$ for the ground state $S$ and $P$-wave heavy mesons and the works on other strong couplings are rare \cite{Huang2010}. The ratios of the strong decay rates are analyzed in the following subsections, which classified the strange charmed mesons according to their total spin and parity. 

\subsection{$D_{s0}^*(2317)$ and $D_{s1}(2460)$} 

\noindent The observed masses of $D_{s0}^*(2317)$ and $D_{s1}(2460)$ states are below $DK$ and $D^*K$ thresholds, respectively. The $BABAR$ \cite{Aubert2003,Aubert2006} observed $D_{s0}^*(2317)$ and $D_{s1}(2460)$ into $D_s^{\pm}\pi^0$ and $D_s^{*\pm}\pi^0$ mode, respectively, or radiatively. Such decay modes are called isospin-breaking modes \cite{Colangelo2003,Colangelo2005}. 

If the mass difference of the parent heavy-light meson with strangeness and daughter nonstrange meson is smaller than the kaon mass, then the parent heavy-light meson with strangeness decays into a heavy-light meson with strangeness and a neutral pion through $\pi^0$-$\eta$ mixing. The violation of isospin symmetry for pseudoscalar mesons within QCD is usually estimated as an admixture parameter $\epsilon$  of the flavor-octet $\eta$ state to the $\pi$. The $\epsilon$ has been related to the current quark masses by Gasser, Treiman and Leutwyler \cite{Gross1979} as:
\begin{equation}
\label{eq:23}
\epsilon = \frac{\sqrt{3}}{4} \frac{m_d - m_u}{m_s - (m_u + m_d)/2} \sim 10^{-2}.
\end{equation}  
\noindent As a result, the decay width formula is multiplied by the factor ${\epsilon}^2 (\sim 10^{-4})$ \cite{Matsuki2012}. For $D_{s0}^*(2317)^{\pm}$ as 1$^3P_0$, our calculated strong decay rate into $D_s^{\pm}\pi^0$ mode is 128.19$g_s^2$ $\times$ 10$^{-4}$ MeV considering the suppression factor originating from the breakdown of isospin symmetry. And, for $D_{s1}(2460)$ as $1P_1$ it is 128.36$g_s^2$ $\times$ 10$^{-4}$ MeV obtained in $D_s^{*\pm}\pi^0$ decay mode.

Experimentally, $BABAR$ \cite{Aubert2006} measured the decay width $< 3.8$ MeV of $D_{s0}^*(2317)^{\pm}$, and from PDG \cite{Tanabashi2018-19,Zyla2020} its branching fraction into $D_s^{*\pm}\pi^0$ is $100^{+0}_{-20}$\% of the total decay rate. So the coupling $g_S$ can evaluate directly by equating our result with experimental measurement. Hence, we write 128.19$g_S^2$ $\times$ 10$^{-4}$ MeV $<$ 3.8 MeV, gives

\begin{equation} 
\label{eq:24}
g_S \le 17.22.  
\end{equation} 

Many theoretical studies \cite{Kher2017,Godfrey2016,Green2017,Devlani2011,Radford2009,Close2005} have computed the radiative electric dipole (E1) transition width of $~~~~$ $D_{s0}^*(2317)^{\pm}$, which is determined in the range of 0.6 KeV to 9 KeV. Therefore, the ratio

\begin{equation} 
\label{eq:25} 
\frac{{\cal{B}}(D_{s0}^*(2317)^{\pm} \rightarrow D_{s}^*(2112)^{\pm}\gamma)}{{\cal{B}}(D_{s0}^*(2317)^{\pm} \rightarrow D_s^{\pm}\pi^0)} \le 2.37\times10^{-3}, 
\end{equation} 

\noindent which is under the limit of CLEO \cite{Besson2003} measurement $<$ 0.059, and measurements of the Belle \cite{Mikami2004} and $BABAR$ \cite{Aubert2006} Collaborations $<$ 0.18 and $<$ 0.16, respectively. For $D_{s1}(2460)^{\pm}$ as 1$P_1$, its strong decay rate into $D_s^{*\pm}\pi^0$ mode is 128.36$g_S^2$ $\times$ 10$^{-4}$ MeV. Using $g_S$ $\le$ 17.22, we get the strong decay width $<$ 3.8 MeV, which is compatible with the $~~~~~$ $BABAR$ measurement $<$ 3.5 MeV \cite{Aubert2006}. Moreover, V. Kher et al. computed its magnetic dipole (M1) transition width 4.9 $\times$ $10^{-2}$ KeV into $D_{s0}^*(2317)^{\pm}$ mode \cite{Kher2017}. So   

\begin{equation} 
\label{eq:26} 
\frac{{\cal{B}}(D_{s1}(2460)^{\pm} \rightarrow D_{s0}^*(2317)^{\pm}\gamma)}{{\cal{B}}(D_{s1}(2460)^{\pm} \rightarrow D_s^{*\pm}\pi^0)} \le 1.29\times10^{-5}, 
\end{equation}

\noindent and is under the limit of $BABAR$ \cite{Aubert2004} result $<$ 0.22. Therefore, from the strong decay analysis and the theoretical predictions of the E1 and M1 transition width, we believe that $D_{s0}^*(2317)^{\pm}$ and $D_{s1}(2460)^{\pm}$ are the members of charmed strange mesons family. Hence, 

\begin{equation}
\label{eq:27} 
\big(D_{s0}^*(2317), D_{s1}(2460)\big) = (0^+, 1^+)_{{\frac{1}{2}}^+} = \big(1^3P_0, 1P_1\big).
\end{equation} 

The $D_{s0}^*(2317)$ and $D_{s1}(2460)$ are still categorized into conventional $P$-wave charmed strange meson family \cite{Kato2019,Chen2017}. Since the masses of $D_{s0}^*(2317)$ and $D_{s1}(2460)$ are below the $DK$ and $D^*K$ thresholds, respectively. Such a low mass puzzle is strongly affected by the coupled channel effect, which is an important nonperturbative QCD effect.  

\subsection{$D_{s1}(2536)$ and $D_{s2}^*(2573)$}

The ratio among the strong decay rates avoided the unknown hadronic coupling and compared directly with experimental observations where available. For $D_{s1}(2536)$ as $1P_1^{\prime}$, the ratio 

\begin{equation} 
\label{eq:28} 
\frac{{\cal{B}}(D_{s1}(2536)^{\pm} \rightarrow D^{*}(2007)^0K^{\pm})}{{\cal{B}}(D_{s1}(2536)^{\pm} \rightarrow D^{*}(2010)^{\pm}K^0)} >1, 
\end{equation} 

\noindent is calculated using the results of Table \ref{tab3} and is consistent with the PDG world average value of 1.18 $\pm$ 0.06 and other experimental observations \cite{Tanabashi2018-19,Zyla2020}. For its spin partner $D_{s2}^*(2573)$, the ratio is 

\begin{equation} 
\label{eq:29} 
\frac{{\cal{B}}(D_{s2}^*(2573)^{\pm} \rightarrow D^*(2007)^0K^{\pm})}{{\cal{B}}(D_{s2}^*(2573)^{\pm} \rightarrow D^0K^{\pm})} \approx 0.087, 
\end{equation}

\noindent calculated for BESIII results listed in Table \ref{tab3} and it is in agreement with the argument $<$ 0.33 of CLEO experiment \cite{Kubota1994}. Also, consistent with the theoretical predictions $\sim$ 0.06 of Refs. \cite{Godfrey2016,Zhong2008} and 0.106 of Ref. \cite{Song2015}. This ratio is close to the 0.076 predicted in Ref. \cite{Matsuki2012}. Hence, $D_{s1}(2536)$ and $D_{s2}^*(2573)$ are well established as a $1P$ state with $\vec{s}_l^P = {\frac{3}{2}}^+$. We write 

\begin{equation}
\label{eq:30} 
\big(D_{s1}(2536), D_{s2}^*(2573)\big) = (1^+, 2^+)_{{\frac{3}{2}}^+} = \big(1P_1^{\prime}, 1^3P_2\big).
\end{equation}

The strong coupling $g_T$ $\sim$ 0.4 is estimated by comparing the sums of the predicted strong decay rates listed in Table 3 with the experimental measurement of the total decay width of $D_{s2}^*(2573)$ $D_s(1^3P_2)$ resonance. We can predict the strong decay rates of $D_{s2}^*(2573)$ meson into $DK$ and $D^*K$ modes as 8.11 MeV and 0.71 MeV, respectively. That is consistent with the results 9.40 MeV $(DK)$ and 0.545 MeV $(D^*K)$ of Ref. \cite{Godfrey2016}, 5.42 MeV $(DK)$ and 0.57 MeV $(D^*K)$ of Ref. \cite{Song2015}, and 3.4 MeV $(DK)$ and 0.27 MeV $(D^*K)$ of Ref. \cite{Matsuki2012}. However, such a comparison is not always justified as especially heavier excited charmed meson resonances can decay to other final states, including those involving higher light pseudoscalar multiplicities, light vector mesons as well as lower lying excited charmed mesons. Such additional decay modes are not included in the effective heavy meson chiral Lagrangians presented in Eqs. (\ref{eq:11}) to (\ref{eq:17}).

\subsection{$D_{s1}^*(2700)$}

After the discovery of comprised two overlapping states $D_{s1}^*(2860)^{\pm}$ and $D_{s3}^*(2860)^{\pm}$ with total spin $1^-$ and $3^-$ respectively by LHCb \cite{Aubert2006}, many experimental and theoretical studies are argued that $D_{s1}^*(2700)^{\pm}$ is the $2^3S_1$ state and the states $D_{s1}^*(2860)^{\pm}$ and $D_{s3}^*(2860)^{\pm}$ are of $1^3D_1$ and $1^3D_3$, respectively \cite{Aaij2014,Song2015,AaijPRD2014,SongChen2015,Godfrey2014/15}. Refs. \cite{Zhong2010,Li2010,Close2007,Li2007} were assigned the $D_{s1}^*(2700)$ as mixing of $2^3S_1$ and $1^3D_1$ states. A molecular state explanation of $D_{s1}^*(2700)$ based on potential model was studied in Ref. \cite{Vinodkumar2007}. For $D_{s1}^*(2700)$ as 2$^3S_1$, the ratio

\begin{equation} 
\label{eq:31} 
\frac{{\cal{B}}(D_{s1}^*(2700)^{\pm} \rightarrow D^{*0}K^{\pm})}{{\cal{B}}(D_{s1}^*(2700)^{\pm} \rightarrow D^0K^{\pm})} \approx 0.9, 
\end{equation}

\noindent is calculated using the results of Table \ref{tab3} and agrees with the measurements 0.88 $\pm$ 0.14 $\pm$ 0.14 and 0.91 $\pm$ 0.13 $\pm$ 0.12 of \cite{Aubert2009}. Therefore, the state $D_{s1}^*(2700)^{\pm}$ with 2$^3S_1$ is dominant in $DK$ mode and is accessible with the experimental measurements. So we write, 

\begin{equation}
\label{eq:32} 
D_{s1}^*(2700)^{\pm} = (1^-)_{{\frac{1}{2}}^-} = \big(2^3S_1\big).
\end{equation}

Its spin partner $D_s(2^1S_0)$ is still not observed experimentally. In theory, it has a mass 2664 MeV covers the Refs. \cite{Kher2017,Godfrey2016,Song2015,Li2011,Ebert2010,DiPierro2001,Lahde2000}. Treating the sum of the partial decay widths as a total decay width (listed in Table \ref{tab3} for unnatural parity states) of $D_s(2^1S_0)$, we obtain the branching fractions $\sim$ 50.7\%, $\sim$ 49.1\%, $\sim$ 5.44 $\times$ 10$^{-3}$\%, and $\sim$ 0.13\% for $D^{*0}K^{\pm}$, $D^{*\pm}K^0$, $D_s^{*\pm}\pi^0$, and $D^{*\pm}\eta$ modes, respectively. The numerical result indicates $D^{*0}K^{\pm}$ and $D^{*\pm}K^0$ are the fundamental decay modes of $D_s(2^1S_0)$. This prediction can be tested by future experimental studies. 

\subsection{$D_{s1}^*(2860)$ and $D_{s3}^*(2860)$}

For $D_{s1}^*(2860)^{\pm}$ as 1$^3D_1$, the ratio

\begin{equation}
\label{eq:33} 
\frac{{\cal{B}}(D_{s1}^*(2860)^{\pm} \rightarrow D^{*0}K^{\pm})}{{\cal{B}}(D_{s1}^*(2860)^{\pm} \rightarrow D^0K^{\pm})} \approx 0.24,
\end{equation}

\noindent is compatible with 0.16 of B. Zhang et al. using $^3P_0$ model \cite{Zhang2007}. Q. T. Song et al. \cite{Song2015} also used $^3P_0$ model and determined its range 0.46 $\sim$ 0.70. The relativistic quark model by Godfrey and Jardine obtained the value of 0.34 \cite{Godfrey2014}. For $D_{s3}^*(2860)^{\pm}$ as 1$^3D_3$, the ratio 

\begin{equation}
\label{eq:34} 
\frac{{\cal{B}}(D_{s3}^*(2860)^{\pm} \rightarrow D^{*0}K^{\pm})}{{\cal{B}}(D_{s3}^*(2860)^{\pm} \rightarrow D^0K^{\pm})} \approx 0.39,
\end{equation}

\noindent is in good agreement with the theoretical results $\sim$ 0.4 of Refs. \cite{Zhong2008,Zhong2010} and consistent with 0.6 of Ref. \cite{Godfrey2016}. Considering $D_{s1}^*(2860)^{\pm}$ as 1$^3D_1$ and $D_{s3}^*(2860)^{\pm}$ as 1$^3D_3$, our calculated branching ratios are much smaller than the $BABAR$ measurement 1.10 $\pm$ 0.15 $\pm$ 0.19 (see in Eq. (\ref{eq:2})) \cite{Aubert2009}. Such a large ratio indicates $D_{sJ}^*(2860)$ consists of at least four resonances $D_{s1}^*(2860)$, $D_{s2}^*(2860)$, $D_{s2}^{*\prime}(2860)$, and $D_{s3}^*(2860)$. The states with total spin-2 near $\approx$ 2.86 GeV mass region still not found experimentally. Our strong decays analysis of $D_{s1}^*(2860)^{\pm}$ and $D_{s3}^*(2860)^{\pm}$, can be confronted with future experimental data.

\subsection{$D_{sJ}(3040)$}

The $D_{sJ}(3040)$ has only been observed in $D^*K$ mode that implies it is of unnatural parity state. The predicted masses for the 2$P_1$ and 2$P_1^{\prime}$ states are nearer to the observed mass (see in Table \ref{tab2}). In Ref. \cite{Godfrey2016}, Godfrey and Moats conclude that 43\% of 2$P_1$ state decay to $D^*K$ final state to the total decay rates and for 2$P_1^{\prime}$ it is 35\%. Mass of a mixed state $(M_{D_{sJ}(3040)^{\pm}}$ is expressed in terms of masses of the two mixing states $(M_{D_{s}(2^1P_1)}$ and $M_{D_{s}(2^3P_1)})$ as

\begin{equation}
\label{eq:35} 
M_{D_{sJ}(3040)^{\pm}} = \lvert a^2 \rvert M_{D_{s}(2P_1)} + \left(1 - \lvert a^2 \rvert \right) M_{D_{s}(2P_1^{\prime})}, 
\end{equation}

\noindent where $\lvert a^2 \rvert = cos^2\theta$ and $\theta$ is the mixing angle. With the help of this equation, we can obtain a mixed state configuration and mixing angle \cite{Shah2012}. Using the calculated masses 3018 MeV and 3038 MeV for $2P_1$ and $2P_1^{\prime}$, respectively, from \cite{Godfrey2016}; we calculate the mixing angle $\theta = 56.79^{\circ}$ corresponding to 54.77\% of the $2P_1$ state and 45.23\% of the $2P_1^{\prime}$ state, i.e. the probability of having 2$P_1$ state is 54.77\% and 2$P_1^{\prime}$ is 45.23\%. Thus, $D_{sJ}(3040)$ is reasonable to interpret as 2$P_1$ state (or the mixing of 2$P_1$ and 2$P_1^{\prime}$ states, but 2$P_1$ is favorable). This state has been studied by many theoretical groups \cite{Li2011,DiPierro2001,Colangelo2010,Zhong2010,Sun2009,Badalian2011,Chen2009}. Di Pierro and Eichten \cite{DiPierro2001} predicted the total decay width $\Gamma \approx$ 210 MeV of 2$P_1$ and is close to the experimental measurement of $D_{sJ}(3040)$ \cite{Aubert2009}, and for the 2$P_1^{\prime}$ state $\Gamma \approx$ 51 MeV. Colangelo and De Fazio \cite{Colangelo2010} argued that for $D_{sJ}(3040)$ the decay width of 2$P_1$ state is more compatible rather than 2$P_1^{\prime}$. Sun and Liu \cite{Sun2009} categorized $D_{sJ}(3040)$ as a $1^+$ state belonging to the $(0^+,1^+)$ doublet. More experimental statistics are needed to clear out the image of $D_{sJ}^*(2860)$ and $D_{sJ}(3040)$ mesons especially.

\section{Regge trajectories}
\label{sec4}

\begin{table}
\caption{\label{tab4}Quantum number assignment of experimentally observed excited strange charmed mesons through strong decays analysis.}
\begin{ruledtabular}
\begin{tabular}{ccccccccccccc}
Exp. \cite{Tanabashi2018-19} (in GeV) & $J^P$ & $\cal{N}$ $^{2S+1}L_J$ \\
\noalign{\smallskip}\hline\noalign{\smallskip} 
2.318 $D_{s0}^*(2317)^{\pm}$ & $0^+$ & $1^3P_0$\\
\noalign{\smallskip}
2.460 $D_{s1}(2460)^{\pm}$ & $1^+$ & $1P_1$\\
\noalign{\smallskip}
2.535 $D_{s1}(2536)^{\pm}$ & $1^+$ & $1P_1^{\prime}$\\
\noalign{\smallskip}
2.569 $D_{s2}^*(2573)^{\pm}$ & $2^+$ & $1^3P_2$\\
\noalign{\smallskip}
2.708 $D_{s1}^*(2700)^{\pm}$ & $1^-$ & $2^3S_1$ \\
\noalign{\smallskip}
2.859 $D_{s1}^*(2860)^{\pm}$ \cite{Aaij2014} & $1^-$ & $1^3D_1$\\ 
\noalign{\smallskip}
2.860 $D_{s3}^*(2860)^{\pm}$ \cite{Aaij2014} & $3^-$ & $1^3D_3$\\ 
\noalign{\smallskip}
3.044 $D_{sJ}(3040)^{\pm}$ \cite{Aubert2009} & $1^+$ & $2P_1$\\ 
\noalign{\smallskip}
\end{tabular}
\end{ruledtabular}
\end{table} 

\begin{figure}
\includegraphics[width=0.49\textwidth]{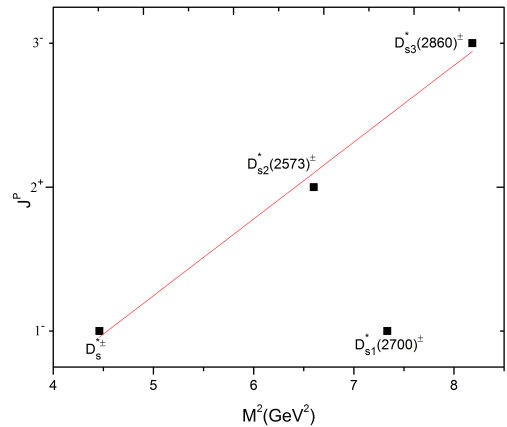}
\includegraphics[width=0.49\textwidth]{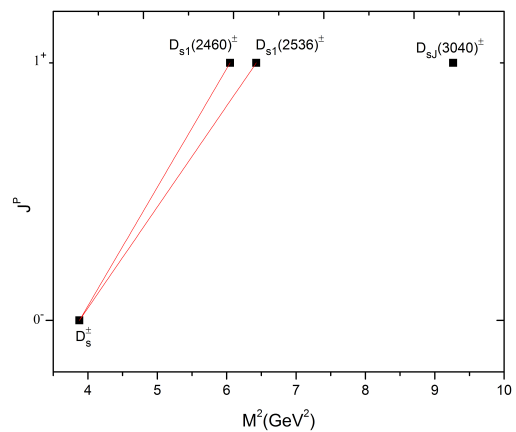}
\caption{\label{fig1}Regge trajectories of strange charmed mesons in $(M^2, J)$ plane with natural parity (upper) and unnatural parity (lower).}
\end{figure}

\begin{figure}
\includegraphics[width=0.49\textwidth]{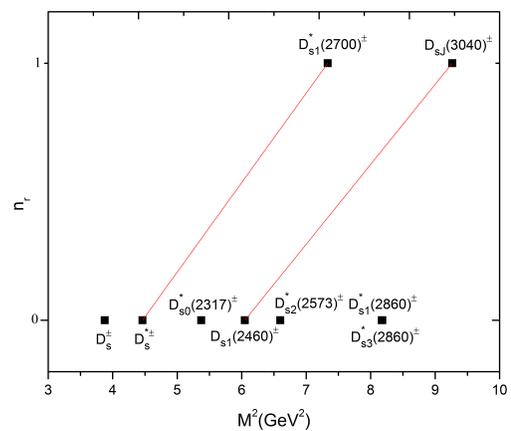}
\caption{\label{fig2}Regge trajectories of strange charmed mesons in $(M^2, n_r)$ plane.}
\end{figure}

One of the most important features of Regge theory is the Regge trajectory by which the mass and the spin of a hadron is related \cite{Regge1959-60}. Our tentative assignment of spin-parity of the excited strange charmed mesons are listed in Table \ref{tab4}, including $D_{s1}^*(2860)^{\pm}$ as 1$^3D_1$ and $D_{s3}^*(2860)^{\pm}$ as 1$^3D_3$, with their respective PDG \cite{Tanabashi2018-19,Zyla2020} world average masses. Using the data from Table \ref{tab4}, we construct the Regge trajectories in $(M^2, J)$ and $(M^2, n_r)$ planes by using the following equations: 
\begin{enumerate}[label=\Roman*.]
\item the Regge trajectory in $(M^2, J)$ plane,
\begin{equation}
\label{eq:36} 
J = \alpha M^2 + \alpha_0;
\end{equation}
\item and the Regge trajectory in $(M^2, n_r)$ plane,
\begin{equation}
\label{eq:37} 
n_r = \beta M^2 + \beta_0;
\end{equation}
\end{enumerate}

\noindent where $\alpha$ and $\beta$ are slopes, and $\alpha_0$ and $\beta_0$ are intercepts. $J$ is the total spin, $n_r (= n -1) = 0, 1, 2, ...$ is the radial principal quantum number, and $M^2$ is the square mass of the $D_s$ mesons. In Fig. \ref{fig1}, we plot the Regge trajectories in $(M^2, J)$ plane with natural $(0^+, 1^-, 2^+, 3^-, ...)$ and unnatural parity $(0^-, 1^+, 2^-, 3^+, ...)$ of the total spin $J$, and the Regge trajectories in $(M^2, n_r)$ plane is presented in Fig. \ref{fig2}. The $D_{s1}(2460)$ and $D_{s1}(2536)$ are identified with $J^P = 1^+$ of the $P$-wave doublets $J^P_{s_l} = (0^+, 1^+)_{{\frac{1}{2}}^+}$ and $J^P_{s_l} = (1^+, 2^+)_{{\frac{3}{2}}^+}$, respectively. Such states $D_{s1}(2460)$ and $D_{s1}(2536)$ having a same origin on the $(M^2, J)$ Regge plane (see the lower panel of Fig. \ref{fig1}).

\begin{table}
\caption{\label{tab5}Fitted parameters of the parent and daughter Regge trajectories in $(M^2, J)$ plane with natural and unnatural parity.}
\begin{ruledtabular}
\begin{tabular}{ccccccccccccc}
& $\alpha$ (GeV$^{-2}$) & $\alpha_0$ & $\alpha$ (GeV$^{-2}$) & $\alpha_0$ &\\ 
\noalign{\smallskip}
\cline{2-3}
\cline{4-5}
\noalign{\smallskip}
Parent & 0.46753 & -1.08583 & 0.46026 & -1.78442 \\
Daughter & $-$ & $-$ & 0.39211 & -1.52018 \\
\end{tabular}
\end{ruledtabular}
\end{table}

\begin{table}
\caption{\label{tab6}Fitted parameters of the Regge trajectories in $(M^2, n_r)$ plane.}
\begin{ruledtabular}
\begin{tabular}{ccccccccccccc}
Meson & $\beta$ (GeV$^{-2}$) & $\beta_0$\\
\noalign{\smallskip}\hline\noalign{\smallskip}
$D_s^{*\pm}$ & 0.34801 & -1.5526 \\
$D_{s1}(2460)^{\pm}$ & 0.31092 & -1.88093 \\
\end{tabular}
\end{ruledtabular}
\end{table}

\begin{table}
\caption{\label{tab7}The masses of strange charmed mesons (in GeV) lying on the $1^3S_1$ and $1^1S_0$ Regge lines in $(M^2, J)$ plane. The masses from PDG \cite{Tanabashi2018-19,Zyla2020} are taken as input.}
\begin{ruledtabular}
\begin{tabular}{ccccccccccccc}
State & $1^3S_1$ & $1^3P_2$ & $1^3D_3$ & $1^3F_4$\\
\noalign{\smallskip}\hline\noalign{\smallskip} 
Present & 2.112 \cite{Tanabashi2018-19,Zyla2020} & 2.569 \cite{Tanabashi2018-19,Zyla2020} & 2.956 & 3.298 \\
Ref. \cite{Godfrey2016} & 2.129 & 2.529 & 2.917 & 3.113  \\
Ref. \cite{Song2015} & 2.115 & 2.571 & 2.883 & 3.143 \\
Ref. \cite{Ebert2010} & 2.111 & 2.571 & 2.971 & 3.300  \\
Ref. \cite{DiPierro2001} & 2.113 & 2.581 & 2.925 & 3.220 \\
\noalign{\smallskip}\hline\noalign{\smallskip}
State & $1^1S_0$ & $1P_1$ & $1D_2$ & $1F_3$\\
\noalign{\smallskip}\hline\noalign{\smallskip}  
Present & 1.969 \cite{Tanabashi2018-19,Zyla2020} & 2.460 \cite{Tanabashi2018-19,Zyla2020} & 2.867 & 3.224 \\
Ref. \cite{Godfrey2016} & 1.979 & 2.549 & 2.900 & 3.108 \\ 
Ref. \cite{Song2015} & 1.967 & 2.529 & $-$ & $-$ \\
Ref. \cite{Ebert2010} & 1.969 & 2.574 & 2.961 & 3.266 \\
Ref. \cite{DiPierro2001} & 1.965 & 2.605 & 2.953 & 3.247 \\
\noalign{\smallskip}\hline\noalign{\smallskip}
State & $1^1S_0$ & $1P_1^{\prime}$ & $1D_2^{\prime}$ & $1F_3^{\prime}$\\
\noalign{\smallskip}\hline\noalign{\smallskip}
Present & 1.969 \cite{Tanabashi2018-19,Zyla2020} & 2.535 \cite{Tanabashi2018-19,Zyla2020} & 2.996 & 3.395 \\
Ref. \cite{Godfrey2016} & 1.979 & 2.556 & 2.926 & 3.218 \\ 
Ref. \cite{Song2015} & 1.967 & 2.534 & $-$ & $-$ \\
Ref. \cite{Ebert2010} & 1.969 & 2.536 & 2.931 & 3.254  \\
Ref. \cite{DiPierro2001} & 1.965 & 2.535 & 2.900 & 3.203 \\
\end{tabular}
\end{ruledtabular}
\end{table}

\begin{table}
\caption{\label{tab8}The masses of strange charmed mesons (in GeV) lying on the $1^3S_1$ and $1P_1$ Regge lines in $(M^2, n_r)$ plane. The masses from PDG \cite{Tanabashi2018-19,Zyla2020} are taken as input.}
\begin{ruledtabular}
\begin{tabular}{ccccccccccccc}
State & $1^3S_1$ & $2^3S_1$ & $3^3S_1$ \\
\noalign{\smallskip}\hline\noalign{\smallskip} 
Present & 2.112 \cite{Tanabashi2018-19,Zyla2020} & 2.708 \cite{Tanabashi2018-19,Zyla2020} & 3.195 & \\
Ref. \cite{Godfrey2016} & 2.129 & 2.732 & 3.193\\ 
Ref. \cite{Ebert2010} & 2.111 & 2.731 & 3.242  \\
Ref. \cite{DiPierro2001} & 2.113 & 2.806 & 3.345  \\
\noalign{\smallskip}\hline\noalign{\smallskip}
State & $1P_1$ & $2P_1$ & $3P_1$ \\
\noalign{\smallskip}\hline\noalign{\smallskip}
Present & 2.460 \cite{Tanabashi2018-19,Zyla2020} & 3.044 \cite{Tanabashi2018-19,Zyla2020} & 3.533 \\
Ref. \cite{Godfrey2016} & 2.549 & 3.018 & 3.416 \\ 
Ref. \cite{Ebert2010} & 2.574 & 3.154 & 3.618 \\
Ref. \cite{DiPierro2001} & 2.605 & 3.165 & $-$ \\
\end{tabular}
\end{ruledtabular}
\end{table}  

The Regge slopes and the intercepts of the Regge lines are extracted (see in Tables \ref{tab5} and \ref{tab6}), and they are assumed to be same for all strange charmed meson multiplets lying on the single Regge line. Not only for spectroscopy purpose, the slopes and the intercepts of the Regge trajectories also has a fundamental importance in hadron physics \cite{Basdevant1985}. Using these parameters, we estimate the masses of experimentally missing strange charmed mesons ($1^3D_3$, $1^3F_4$, $1D_2$, $1F_3$, $1D_2^{\prime}$, $1F_3^{\prime}$, $3^3S_1$, and $3P_1$) lying on these Regge lines, which are presented in Tables \ref{tab7} and \ref{tab8} with other theoretical predictions. Our results are reasonably close to the predictions of Refs. \cite{Godfrey2016,Song2015,Ebert2010,DiPierro2001}. Their strong decays into light pseudoscalar mesons ($\pi$, $K$, and $\eta$) are calculated, which are presented in Table \ref{tab9} with the ratio, $\hat{\Gamma} = \frac{\Gamma}{\Gamma \big({{\cal{N}}^{2S+1}L_J \rightarrow D^{*0} K^{\pm}}\big)}$, among the decay widths. Only the natural parity states can decay into two-pseudoscalar mesons states. From the strong decay analysis, the states $D_s$(1$D_2$), $D_s$(1$^1F_3$), $D_s$(1$D_2^{\prime}$), $D_s$(1$F_3^{\prime}$), $D_s$(3$^3S_1$), and $D_s$(3$P_1$) dominantly decay in $D^*K$ mode, and the states $D_s$($1^3D_3$) and $D_s$($1^3F_4$) are mainly found in $D^0 K^{\pm}$. That is useful in further experimental search. The Regge trajectories with unnatural parity predicted the masses of the states $D_{s2}^*(2860)^+$ $(1D_2)$ and $D_{s2}^{*\prime}(2860)^+$ $(1D_2^{\prime})$ as 2.867 GeV and 2.996 GeV, respectively. The ratio,

{\small{\begin{equation}
\label{eq:38} 
\frac{{\cal{B}}(D_{s1}^*(2860)^{\pm} \rightarrow D^{*0}K^{\pm}) + {\cal{B}}(D_{s2}^{*}(2860)^{\pm} \rightarrow D^{*0}K^{\pm})}{{\cal{B}}(D_{s1}^*(2860)^{\pm} \rightarrow D^0K^{\pm})} \approx 1.01,
\end{equation}}}

\noindent is calculated using the results of Tables \ref{tab3} and \ref{tab9}, which is consistent with the $BABAR$ measurement 1.10 $\pm$ 0.15 $\pm$ 0.19 \cite{Aubert2009} (see in Eq. (\ref{eq:2})). The coupling $g_X$ $\approx$ 0.20, fit the sum of the partial decay rates of $D_{s1}^*(2860)^{\pm}$ (listed in Table \ref{tab3}) with the experimental decay width 159 $\pm$ 23 $\pm$ 27 MeV \cite{Aaij2014}. The ratio in Eq. (\ref{eq:38}) considering the two contributions, $D_{s1}^*(2860)^{\pm}$ and $D_{s2}^{*}(2860)^{\pm}$. In 2014, the LHCb Collaboration found the overlapping of spin-1 and spin-3 at mass $\bar{D}^0K^-$ $\approx$ 2.86 GeV with a significance of more than 10 standard deviations \cite{Aaij2014}. Wang \cite{Wang2015} conclude that $D_{sJ}^*(2860)$ might consist of at least four resonance states $D_{s1}^*(2860)$, $D_{s2}^*(2860)$, $D_{s2}^{*\prime}(2860)$, $D_{s3}^*(2860)$, and that are required for the large ratio of 1.10 $\pm$ 0.15 $\pm$ 0.19. Using the strong decay rates of $1^3D_1$ from Table \ref{tab3}, and 1$D_2$, 1$D_2^{\prime}$ and $1^3D_3$ from Table \ref{tab9}, we compute the ratio,

\begin{widetext}
{\small{\begin{equation}
\label{eq:39} 
\frac{{\cal{B}}(D_{s1}^*(2860)^{\pm} \rightarrow D^{*0}K^{\pm}) + {\cal{B}}(D_{s2}^{*}(2860)^{\pm} \rightarrow D^{*0}K^{\pm}) + {\cal{B}}(D_{s2}^{*\prime}(2860)^{\pm} \rightarrow D^{*0}K^{\pm}) + {\cal{B}}(D_{s3}^*(2860)^{\pm} \rightarrow D^{*0}K^{\pm})}{{\cal{B}}(D_{s1}^*(2860)^{\pm} \rightarrow D^0K^{\pm}) +{\cal{B}}(D_{s3}^*(2860)^{\pm} \rightarrow D^0K^{\pm})} \approx 1.16,
\end{equation}}}
\end{widetext}  

\noindent that is compatible with the experimental value 1.10 $\pm$ 0.15 $\pm$ 0.19 \cite{Aubert2009}. The couplings $g_X$ $\approx$ 0.20 and $g_Y$ $\approx$ 0.25 are obtained by comparing the sum of the partial decay widths of $D_{s1}^*(2860)^{\pm}$ as $1^3D_1$ (from Table \ref{tab3}) and $D_{s3}^*(2860)^{\pm}$ as $1^3D_3$ (from Table \ref{tab9}) with their respective experimental measurements 159 $\pm$ 23 $\pm$ 27 MeV and 53 $\pm$ 7 $\pm$ 7 MeV \cite{Aaij2014}. Assuming that the sum of these partial decay widths is contributing dominantly to the total decay width. Godfrey and Moats \cite{Godfrey2016} determine the total decay width 197.2 MeV of $1^3D_1$ and 46 MeV of $1^3D_3$, contributing only their decays to ground state charmed mesons plus light pseudoscalar mesons ($\pi$, $\eta$, and $K$), and are consistent with the LHCb \cite{Aaij2014}. Here, we used the mass 2.956 GeV of $D_{s3}^*(2860)^{\pm}$ predicted in Table \ref{tab7}. Experimentally, the mass difference $M_{D_{s1}^*(2860)^{\pm}}-M_{D_{s3}^*(2860)^{\pm}}$ $\approx$ 1.5 MeV \cite{Aaij2014} is very much smaller than the theoretical predictions (see in Table \ref{tab2}). For such a large ratio, 1.10 $\pm$ 0.15 $\pm$ 0.19, the mass of $D_{s3}^*(2860)^{\pm}$ should be expected to higher than 2.86 GeV.

\begin{table}
\caption{\label{tab9}The ratios $\hat{\Gamma} = \frac{\Gamma}{\Gamma \big({{\cal{N}}^{2S+1}L_J \rightarrow D^{*0} K^{\pm}}\big)}$ of the strong decays of the strange charmed mesons lying on the Regge lines.}
\begin{ruledtabular}
\begin{tabular}{ccccccccccccc}
$\cal{N}$ $^{2S+1}L_J$ & Decay & Decay & Ratio\\
& mode & width & $\hat{\Gamma}$ \\
\noalign{\smallskip}\hline\noalign{\smallskip} 
1$^3D_3$ & $D^{*0} K^{\pm}$ & 127.02$g_Y^2$ & 1 \\
& $D^{*\pm} K^0$ & 120.87$g_Y^2$ & 0.95 \\
& $D{_{s}^{*\pm}} \pi^0$ & 96.65$g_Y^2$ $\times$ 10$^{-4}$ & 0.76 $\times$ 10$^{-4}$\\
& $D{_{s}^{*\pm}} \eta$ & 20.01$g_Y^2$ & 0.16 \\
& $D^0 K^{\pm}$ & 258.78$g_Y^2$ & 2.04 \\
& $D^{\pm} K^0$ & 251.54$g_Y^2$ & 1.98 \\
& $D{_{s}^{\pm}} \pi^0$ & 169.72$g_Y^2$ $\times$ 10$^{-4}$ & 1.34 $\times$ 10$^{-4}$ \\
& $D{_{s}^{\pm}} \eta$ & 63.97$g_Y^2$ & 0.5 \\
\noalign{\smallskip}\hline\noalign{\smallskip} 
1$^3F_4$ & $D^{*0} K^{\pm}$ & 4333.97$g_R^2$ & 1 \\
& $D^{*\pm} K^0$ & 4191.62$g_R^2$ & 0.97 \\
& $D{_{s}^{*\pm}} \pi^0$ & 2519.92$g_R^2$ $\times$ 10$^{-4}$ & 0.58 $\times$ 10$^{-4}$ \\
& $D{_{s}^{*\pm}} \eta$ & 1158.24$g_R^2$ & 0.27 \\
& $D^0 K^{\pm}$ & 7428.53$g_R^2$ & 1.71 \\
& $D^{\pm} K^0$ & 7262.69$g_R^2$ & 1.68\\
& $D{_{s}^{\pm}} \pi^0$ & 4167.22$g_R^2$ $\times$ 10$^{-4}$ & 0.96 $\times$ 10$^{-4}$ \\
& $D{_{s}^{\pm}} \eta$ & 2405.38$g_R^2$ & 0.56 \\
\noalign{\smallskip}\hline\noalign{\smallskip} 
1$D_2$ & $D^{*0} K^{\pm}$ & 1111.70$g_X^2$ & 1 \\
& $D^{*\pm} K^0$ & 1078.20$g_X^2$ & 0.97 \\
& $D{_{s}^{*\pm}} \pi^0$ & 537.35$g_X^2$ $\times$ 10$^{-4}$ & 0.48 $\times$ 10$^{-4}$ \\
& $D{_{s}^{*\pm}} \eta$ & 278.52$g_X^2$ & 0.25 \\
\noalign{\smallskip}\hline\noalign{\smallskip} 
1$F_3$ & $D^{*0} K^{\pm}$ & 1919.35$g_Z^2$ & 1 \\
& $D^{*\pm} K^0$ & 1875.56$g_Z^2$ & 0.98 \\
& $D{_{s}^{*\pm}} \pi^0$ & 886.78$g_Z^2$ $\times$ 10$^{-4}$ & 0.46 $\times$ 10$^{-4}$ \\
& $D{_{s}^{*\pm}} \eta$ & 655.12$g_Z^2$ & 0.34 \\
\noalign{\smallskip}\hline\noalign{\smallskip} 
1$D_2^{\prime}$ & $D^{*0} K^{\pm}$ & 311.60$g_Y^2$ & 1 \\
& $D^{*\pm} K^0$ & 297.87$g_Y^2$ & 0.96 \\
& $D{_{s}^{*\pm}} \pi^0$ & 223.45$g_Y^2$ $\times$ 10$^{-4}$ & 0.72 $\times$ 10$^{-4}$ \\
& $D{_{s}^{*\pm}} \eta$ & 57.06$g_Y^2$ & 0.18 \\
\noalign{\smallskip}\hline\noalign{\smallskip} 
1$F_3^{\prime}$ & $D^{*0} K^{\pm}$ & 14624.55$g_R^2$ & 1 \\
& $D^{*\pm} K^0$ & 14206.14$g_R^2$ & 0.97 \\
& $D{_{s}^{*\pm}} \pi^0$ & 8146.97$g_R^2$ $\times$ 10$^{-4}$ & 0.55 $\times$ 10$^{-4}$ \\
& $D{_{s}^{*\pm}} \eta$ & 4424.50$g_R^2$ & 0.30 \\
\noalign{\smallskip}\hline\noalign{\smallskip} 
3$^3S_1$ & $D^{*0} K^{\pm}$ & 2641.09$g_H^2$ & 1 \\
& $D^{*\pm} K^0$ & 2609.54$g_H^2$ & 0.99 \\
& $D{_{s}^{*\pm}} \pi^0$ & 1468.44$g_H^2$ $\times$ 10$^{-4}$ & 0.56 $\times$ 10$^{-4}$ \\
& $D{_{s}^{*\pm}} \eta$ & 1267.05$g_H^2$ & 0.48 \\
& $D^0 K^{\pm}$ & 1684.57$g_H^2$ & 0.64 \\
& $D^{\pm} K^0$ & 1672.93$g_H^2$ & 0.63 \\
& $D{_{s}^{\pm}} \pi^0$ & 919.94$g_H^2$ $\times$ 10$^{-4}$ & 0.35 $\times$ 10$^{-4}$ \\
& $D{_{s}^{\pm}} \eta$ & 880.24$g_H^2$ & 0.33 \\
\noalign{\smallskip}\hline\noalign{\smallskip} 
3$P_1$ & $D^{*0} K^{\pm}$ & 9113.52$g_S^2$ & 1 \\
& $D^{*\pm} K^0$ & 9082.23$g_S^2$ & 0.99 \\
& $D{_{s}^{*\pm}} \pi^0$ & 4103.49$g_S^2$ $\times$ 10$^{-4}$ & 0.45 $\times$ 10$^{-4}$ \\
& $D{_{s}^{*\pm}} \eta$ & 5408.97$g_S^2$ & 0.59 \\
\end{tabular}
\end{ruledtabular}
\end{table}  

\section{Summary}
\label{sec5}

In this article, the excited strange charmed mesons $~~~~$ $D_{s0}^*(2317)^{\pm}$, $D_{s1}(2460)^{\pm}$, $D_{s1}(2536)^{\pm}$, $D_{s2}^*(2573)^{\pm}$, $~~~~$ $D_{s1}^*(2700)^{\pm}$, $D_{s1}^*(2860)^{\pm}$, $D_{s3}^*(2860)^{\pm}$, and $D_{sJ}^*(3040)^{\pm}$ (observed by the various experimental groups BESIII \cite{Ablikim2019}, LHCb \cite{Aaij2014,Aaij2012}, and $BABAR$ \cite{Aubert2009,Aubert2006,Aubert2006_2}) are examined as 1$^3P_0$, 1$P_1$, 1$P_1^{\prime}$, 1$^3P_2$, 2$^3S_1$, 1$^3D_1$, 1$^3D_3$, and 2$P_1$ or 2$P_1^{\prime}$, respectively according to their masses, spin and parity. Then we apply HQET to calculate their strong decays into ground state charmed mesons along with light pseudoscalar mesons. The strong decay rates and the branching ratio among decay widths confirmed the states $~~~~$ $D_{s0}^*(2317)$ and $D_{s1}(2460)$ are belonging to a doublet $~~~~$ $(1^3P_0, 1P_1)$, and the states $D_{s1}(2536)$ and $D_{s2}^*(2573)$ are from $(1P_1^{\prime}, 1^3P_2)$. The $D_{s1}^*(2700)$ is confirmed as a first excited state of $S$-wave with $J^P = 1^+$. The decay behavior of $D_s$($2^1S_0$) is predicted, which is still missing in the experiment. It is suitable to search in $D_s^{*}K$ decay modes. The calculated branching ratio $\frac{D^*K}{DK}$ for $D_{s1}^*(2860)$ with 1$^3D_1$ is $\approx$ 0.24, much smaller than the $BABAR$ measurement of 1.10 $\pm$ 0.15 $\pm$ 0.19 \cite{Aubert2009}. After the observation of LHCb Collaboration \cite{Aaij2014}, overlapping of spin-1 and spin-3 components (i.e. the states $D_{s1}^*(2860)$ and $D_{s3}^*(2860)$ respectively), Wang \cite{Wang2015} argued that the large ratio 1.10 $\pm$ 0.15 $\pm$ 0.19 requires $D_{sJ}^*(2860)$ consists of at least four resonance states $~~~~$ $D_{s1}^*(2860)$, $D_{s2}^*(2860)$, $D_{s2}^{*\prime}(2860)$, $D_{s3}^*(2860)$. Experimentally, the $D_s$ meson of total spin-2 with unnatural parity is not observed yet. Using the quantum number assignments of $D_{s0}^*(2317)^{\pm}$, $D_{s1}(2460)^{\pm}$, $D_{s1}(2536)^{\pm}$, $D_{s2}^*(2573)^{\pm}$, $~~~~$ $D_{s1}^*(2700)^{\pm}$ as $1^3P_0$, $1P_1$, $1P_1^{\prime}$, $1^3P_2$, and $2^3S_1$ respectively from Table \ref{tab4}, and states $D_{s1}^*(2860)^{\pm}$ and $D_{s3}^*(2860)^{\pm}$ with  $J^P = 1^-$ and $3^-$ of the $1D$ family from PDG \cite{Tanabashi2018-19,Zyla2020}, we construct the Regge trajectories in $(M^2, J)$ plane with natural and unnatural parity of the total spin. Fixing the slopes and the intercepts of the straight Regge lines in $(M^2, J)$ plane, we estimate the masses of $1^3D_3$, $1^3F_4$, $1D_2$, $1F_3$, 1$D_2^{\prime}$, and 1$F_3^{\prime}$ strange charmed mesons. The strong decays of $1^3D_1$ from Table \ref{tab3}, and $1D_2$, 1$D_2^{\prime}$ and $1^3D_3$ from Table \ref{tab9}, we determine the branching ratio $\frac{D^*K}{DK}$ $\approx 1.16$, that is in agreement with 1.10 $\pm$ 0.15 $\pm$ 0.19 \cite{Aubert2009}.

The $D_{sJ}^*(3040)^{\pm}$ is the good candidate for 2$P_1$ and 2$P_1^{\prime}$ states. From the mixing angel calculation, the $D_{sJ}^*(3040)^{\pm}$ is more appropriate to assign 2$P_1$ quantum state. The $D_{sJ}^*(3040)^{\pm}$ with 2$P_1$ we construct the Regge trajectories in $(M^2, n_r)$ plane. We estimate the masses of $3^3S_1$ and 3$P_1$ strange charmed mesons by fixing the slopes and the intercepts of the Regge lines in $(M^2, n_r)$ plane. The masses of $1^3D_3$ and $1^3F_4$ states are in good agreement with the results of D. Ebert et al. \cite{Ebert2010}, and for $3^3S_1$ state, our prediction 3.195 GeV is very near to Godfrey and Moats result \cite{Godfrey2016}. From the strong decay rates and the ratio $\hat{\Gamma} = \frac{\Gamma}{\Gamma \big({{\cal{N}}^{2S+1}L_J \rightarrow D^{*0} K^{\pm}}\big)}$, we conclude that the states $D_s$($1D_2$), $D_s$($1F_3$), $D_s$(1$D_2^{\prime}$), $D_s$(1$F_3^{\prime}$), $D_s$($3^3S_1$), and $D_s$(3$P_1$) are dominant in $D^*K$ mode, and the states $D_s$($1^3D_3$), $D_s$($1^3F_4$) are dominant in $DK$ mode. Thus these predictions have opened a window to investigate higher excitation of strange charmed mesons at LHCb, BESIII, KEK-B etc., and the future facilities Belle II and $\overline{\mbox{\sffamily P}}${\sffamily ANDA}.

\section*{Acknowledgment}

Keval Gandhi was inspired by the work of Prof. A. V. Manohar, Prof. M. B. Wise, Prof. M. Neubert, Prof. A. F. Falk, Prof. M. E. Luke, Prof. R. Casalbuoni, Prof. S. Campanella, Prof. P. Colangelo, Prof. F. De Fazio and Prof. Z.-G. Wang on HQET and would like to thank them for their valuable contributions to this field.

\end{document}